\documentclass[twocolumn,english,superscriptaddress,aps,prc]{revtex4-2}
\usepackage[T1]{fontenc}
\usepackage[utf8]{inputenc}
\usepackage{color}
\usepackage{babel}
\usepackage{amsmath}
\usepackage{amsthm}
\usepackage{booktabs}
\usepackage{threeparttable}
\usepackage{multirow}
\usepackage{graphicx}
\usepackage{amsmath}

\usepackage{amsmath}
\usepackage[normalem]{ulem}
\usepackage{xcolor}

\newcommand{\redsout}{%
  \bgroup
  \markoverwith{\textcolor{red}{\rule[0.5ex]{2pt}{0.4pt}}}%
  \ULon
}

\usepackage[pdfusetitle,
 bookmarks=true,bookmarksnumbered=false,bookmarksopen=false,
 breaklinks=false,pdfborder={0 0 1},backref=false,colorlinks=true]
 {hyperref}
\hypersetup{
 linkcolor=blue,citecolor=cyan}

\makeatletter

\providecommand{\tabularnewline}{\\}

\usepackage{babel}
\usepackage{amsthm}

\usepackage{xcolor}

\makeatother

\begin{document}
\title{Global polarization of $\Lambda$, $\Xi^{-}$, and $\Omega^{-}$ hyperons in Au+Au collisions \\ 
at RHIC BES-II energies}
\author{Gen-Hui Li}
%\email{genhuili@mail.ustc.edu.cn}

\affiliation{Department of Modern Physics, University of Science and Technology
of China, Anhui 230026, China}
\author{Cong Yi}
\email{congyi@mail.ustc.edu.cn}

\affiliation{Institute of Particle Physics and Key Laboratory of Quark and Lepton
Physics (MOE), Central China Normal University, Wuhan 430079, China}
\author{Xiang-Yu Wu}
%\email{xiangyu.wu2@mail.mcgill.ca}

\affiliation{Department of Physics, McGill University, Montreal, QC, Canada H3A
2T8}
%\affiliation{Department of Physics and Astronomy, Wayne State University, Detroit
%MI 48201}

\author{Shi Pu}
\email{shipu@ustc.edu.cn}

\affiliation{Department of Modern Physics, University of Science and Technology
of China, Anhui 230026, China}
\affiliation{Southern Center for Nuclear-Science Theory (SCNT), Institute of Modern
Physics, Chinese Academy of Sciences, Huizhou 516000, Guangdong Province,
China}

\author{Guang-You Qin}
\email{guangyou.qin@ccnu.edu.cn}

\affiliation{Institute of Particle Physics and Key Laboratory of Quark and Lepton
Physics (MOE), Central China Normal University, Wuhan 430079, China}

\begin{abstract}
We investigate the global spin polarization of $\Lambda$ hyperons and the multi-strange hyperons $\Xi^{-}$ and $\Omega^{-}$ in Au+Au collisions across the RHIC Beam Energy Scan II (BES-II)  energy range, $\sqrt{s_{NN}}=7.7$--$27$ GeV, with feed-down effects included. The polarization is computed using the modified Cooper--Frye formula, which includes contributions from thermal vorticity, the thermal shear tensor, and the gradient of the baryon chemical potential, combined with the (3+1)-dimensional viscous hydrodynamic framework CLVisc with SMASH initial conditions. We present the global polarization as a function of collision energy, centrality, transverse momentum, and rapidity with feed-down effects. We find that the global polarization of $\Omega^{-}$ is systematically larger than those of $\Lambda$ and $\Xi^{-}$ because of its larger spin quantum number, but it remains below the central value of the recent STAR measurement. This discrepancy may suggest that additional mechanisms, such as spin correlations among strange quarks inside the $\Omega^{-}$, could contribute to the observed $\Omega^{-}$ polarization. We also find that the global-polarization splitting between hyperons and anti-hyperons increases toward lower collision energies and is dominated by the chemical-potential-gradient contribution.
\end{abstract}
\maketitle

\section{Introduction}

In noncentral relativistic heavy-ion collisions, the quark-gluon plasma (QGP) created in the overlap region carries enormous orbital angular momentum and behaves as the most vortical fluid observed so far, with an angular velocity reaching up to $10^{22} \; s^{-1}$ \cite{STAR:2017ckg}. Such an immense vorticity can polarize the spins of final-state hadrons along the direction of the initial orbital angular momentum through spin--orbit coupling \cite{Liang:2004ph,Liang:2004xn}. This phenomenon is known as global spin polarization in relativistic heavy-ion collisions. Over the past decade, the global polarization of $\Lambda$ and $\bar{\Lambda}$ hyperons has been extensively measured from a few GeV up to the TeV scale and in different collision systems \cite{STAR:2007ccu,STAR:2021beb,ALICE:2019onw,STAR:2018gyt,STAR:2017ckg,STAR:2023nvo,STAR:2025dgs}. On the theoretical side, numerical calculations based on the spin-polarization vector derived from the quantum statistical model \cite{Becattini:2013fla} or from quantum kinetic theory \cite{Fang:2016vpj} have described the collision-energy dependence of the global polarization of $\Lambda$ hyperons over $\sqrt{s_{NN}}=7.7$--$200$ GeV \citep{Becattini:2007nd,Betz:2007kg,Becattini:2013fla,Becattini:2013vja,Csernai:2013bqa,Fang:2016vpj,Karpenko:2016jyx,Xie:2017upb,Li:2017slc,Sun:2017xhx,Shi:2017wpk,Xia:2018tes,Shi:2019wzi,Fu:2020oxj,Lei:2021mvp,Ambrus:2020oiw,Vitiuk:2019rfv,Gao:2020vbh, Wu:2022mkr,Becattini:2024uha}. At low collision energies, the global polarization of $\Lambda$ hyperons is found to continue increasing as the collision energy decreases \citep{STAR:2021beb,HADES:2022enx}. Although several studies have investigated the effects of vorticity \citep{Deng:2020ygd, Ivanov:2020udj, Deng:2021miw, Yi:2026rbz}, the mechanism underlying the global polarization at such low collision energies is still not fully understood. Very recently, the global polarization of hypernuclei and protons has also been studied \citep{Sun:2025oib,Liu:2025kpp,Zheng:2025ngn,Xu:2026hxz}.

Although significant progress has been made, several issues concerning global polarization remain to be understood. First, most existing studies have focused on the collision-energy dependence of the global polarization, while its dependence on rapidity, centrality, and transverse momentum still requires further investigation, especially in view of future experimental measurements. Second, it has been found that the polarization induced by the shear tensor and by the gradient of the baryon chemical potential \citep{Hidaka:2017auj,Liu:2020dxg,Becattini:2021suc,Liu:2021uhn,Fu:2021pok,Becattini:2021iol,Yi:2021ryh,Ryu:2021lnx,Florkowski:2021xvy,Buzzegoli:2022fxu,Becattini:2022zvf,Palermo:2024tza}, which are closely related to spin Hall-type effects, can affect the polarization of hyperons along the beam direction \citep{STAR:2019erd,STAR:2023eck,ALICE:2021pzu,Voloshin:2017kqp,Liu:2019krs,Wu:2020yiz,Wu:2019eyi,Xia:2019fjf,Becattini:2019ntv,Li:2021jvn,Ivanov:2019ern,Guo:2021udq,Ayala:2021xrn,Deng:2021miw}. Therefore, a natural question is how large the contributions from shear-induced polarization and spin Hall effects are to the global polarization.
Third, the polarization of multi-strange hyperons \cite{STAR:2020xbm} has also attracted considerable attention. These particles provide a useful probe of whether spin correlations \cite{Sheng:2022ffb,Lv:2026kct,Zhang:2024hyq,Yang:2017sdk,Sheng:2019kmk,Xia:2020tyd,Wagner:2022gza,Sheng:2022ssp,Muller:2021hpe,Lv:2024uev,Wei:2023pdf,Dong:2023cng,Kumar:2023ghs,Sheng:2023urn,Chen:2023hnb,Sheng:2022wsy,Chen:2024afy,Li:2022vmb,Sheng:2024kgg,Sheng:2025puj} exist among strange quarks inside a hyperon and, if so, how such correlations affect the observed global polarization. This remains an open issue that calls for systematic theoretical investigation. Finally, the origin of the global-polarization splitting between hyperons and anti-hyperons remains unclear. Besides the splitting induced by electromagnetic fields ~\citep{Muller:2018ibh, Guo:2019joy, Buzzegoli:2022qrr, Xu:2022hql, Peng:2022cya}, gradients of the chemical potential can also contribute to this splitting \cite{Ryu:2021lnx, Wu:2022mkr}. Therefore, it is important to clarify the relative contributions of these different mechanisms.

\begin{figure*}[t]
\centering 
\includegraphics[width=0.95\linewidth]{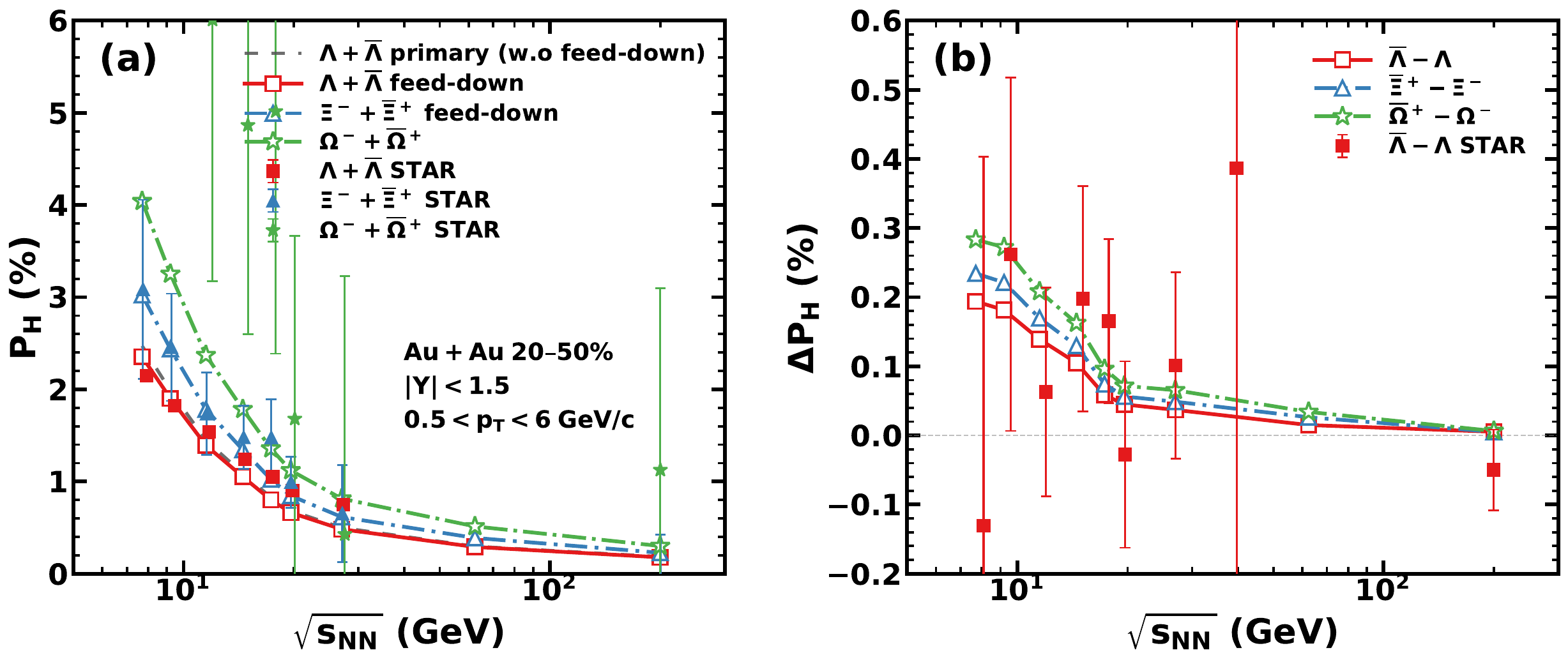} \caption{(a) Global polarization $P_H$ and (b) global-polarization splitting $\Delta P_H$ of $\Lambda$, $\Xi^{-}$, and $\Omega^{-}$ hyperons in $20$--$50\%$ central Au+Au collisions as functions of collision energy. The rapidity and transverse-momentum ranges are set to $|Y|<1.5$ and $0.5<p_T<6$ GeV/c, respectively. Red filled squares, blue filled triangles, and green filled stars represent the experimental data points for the global polarization of $\Lambda$, $\Xi$, and $\Omega$ extracted from Ref.~\cite{Lu:2026ghe}, respectively. Open symbols with the same colors represent the corresponding theoretical results. }
\label{fig:energy} 
\end{figure*}

Interestingly, the high-statistics data from the RHIC 
Beam Energy Scan II (BES-II) \citep{STAR:2002eio} are expected to provide more precise measurements of both the global polarization of multi-strange hyperons. Therefore, it is timely to study the global polarization at the new collision energies covered by BES II.

To address these issues and provide predictions for global polarization in the RHIC BES-II energy range, we employ the (3+1)-dimensional hydrodynamic framework CLVisc \cite{Pang:2012he,Pang:2018zzo,Wu:2021fjf}, with SMASH initial conditions \citep{Weil:2016zrk,Schafer:2019edr,Mohs:2019iee,Hammelmann:2019vwd,Mohs:2020awg,Schafer:2021csj,Inghirami:2022afu}, to compute the global polarization of the multi-strange hyperons $\Xi^-$ and $\Omega^-$, together with the hyperon--anti-hyperon polarization splitting and the corresponding feed-down effects, at $\sqrt{s_{NN}}=7.7$--$27$ GeV.

This paper is organized as follows. In Secs.~\ref{sec:Theoretical-and-numerical} and \ref{sec:numerical}, we introduce the theoretical framework and the numerical setup, respectively.  In Sec.~\ref{sec:feed-down}, we derive the centrality, transverse momentum and rapidity dependent feed-down corrections to global polarization. In Sec.~\ref{sec:Results-and-discussion}, we present our numerical results for the global polarization of $\Lambda$, $\Xi^-$, and $\Omega^-$ hyperons and analyze its dependence on collision energy, centrality, transverse momentum, and rapidity. Finally, we summarize our findings in Sec.~\ref{sec:Summary}.

\section{Theoretical framework}
\label{sec:Theoretical-and-numerical}

In this section, we briefly describe our theoretical framework. Based on the quantum statistical model \citep{Becattini:2013fla} or quantum kinetic theory \citep{Fang:2016vpj}, the mean spin-polarization vector of spin-$s$ fermions, normalized by the particle number and the spin quantum number, is given by
\begin{eqnarray}
\mathcal{P}^{\mu}(\mathbf{p}) & = & \frac{(s+1)\int d\Sigma\cdot p\,\mathcal{J}_{5}^{\mu}(p,x)}{3m_{f}\int d\Sigma\cdot p\,f_{{\rm eq}}},\label{eq:modifed-CF}
\end{eqnarray}
where $m_{f}$ is the fermion mass entering the normalization of the Pauli--Lubanski pseudovector, $p^{\mu}$ is the on-shell four-momentum of the fermion satisfying $p^{2}=m_{f}^{2}$, and $f_{{\rm eq}}={\exp[(p^{\alpha}u_{\alpha}-\mu)/T]+1}^{-1}$ is the Fermi--Dirac distribution with flow velocity $u_{\alpha}$, chemical potential $\mu$, and temperature $T$. The vector $d\Sigma^{\mu}$ denotes the normal vector of the freeze-out hypersurface. The quantity $\mathcal{J}_{5}^{\mu}$ denotes the axial current in phase space, which is proportional to the spin current in phase space. Substituting $\mathcal{J}_{5}^{\mu}$ up to order $\hbar$ and neglecting electromagnetic fields, one can decompose $\mathcal{P}^{\mu}$ as \cite{Yi:2021ryh, Wu:2022mkr},
\begin{eqnarray}
\mathcal{P}^{\mu} & = & \mathcal{P}_{\text{thermal}}^{\mu}+\mathcal{P}_{\text{th-shear}}^{\mu}+\mathcal{P}_{\text{chemical}}^{\mu},
\end{eqnarray}
with 
\begin{eqnarray}
\mathcal{P}_{\text{thermal}}^{\mu}(\mathbf{p}) & = & \frac{s+1}{6m_{f}}\int d\Sigma\cdot\mathcal{N}_{p}\,\epsilon^{\mu\nu\alpha\beta}p_{\nu}\varpi_{\alpha\beta},\nonumber \\
\mathcal{P}_{\text{th-shear}}^{\mu}(\mathbf{p}) & = & \frac{s+1}{3m_{f}}\int d\Sigma\cdot\mathcal{N}_{p}\,\frac{\epsilon^{\mu\nu\alpha\beta}p_{\nu}n_{\beta}}{(n\cdot p)}p^{\sigma}\xi_{\sigma\alpha},\nonumber \\
\mathcal{P}_{\text{chemical}}^{\mu}(\mathbf{p}) & = & \frac{s+1}{3m_{f}}\int d\Sigma\cdot\mathcal{N}_{p}\,\frac{\epsilon^{\mu\nu\alpha\beta}p_{\alpha}}{(u\cdot p)}u_{\beta}\partial_{\nu}\frac{\mu}{T},
\end{eqnarray}
where $\mathcal{N}_{p}^{\mu}=p^{\mu}f_{{\rm eq}}(1-f_{{\rm eq}})/\int d\Sigma\cdot p\,f_{{\rm eq}}$.
The subscripts ``thermal'', ``th-shear'', and ``chemical'' label
the contributions from the thermal vorticity $\varpi_{\alpha\beta}=\frac{1}{2}\left[\partial_{\alpha}\left(u_{\beta}/T\right)-\partial_{\beta}\left(u_{\alpha}/T\right)\right]$,
thermal shear tensor $\xi_{\alpha\beta}=\frac{1}{2}\left[\partial_{\alpha}\left(u_{\beta}/T\right)+\partial_{\beta}\left(u_{\alpha}/T\right)\right]$,
and chemical-potential gradient $\nabla(\mu/T)$, respectively.

\begin{figure*}
\centering \includegraphics[width=0.95\linewidth]{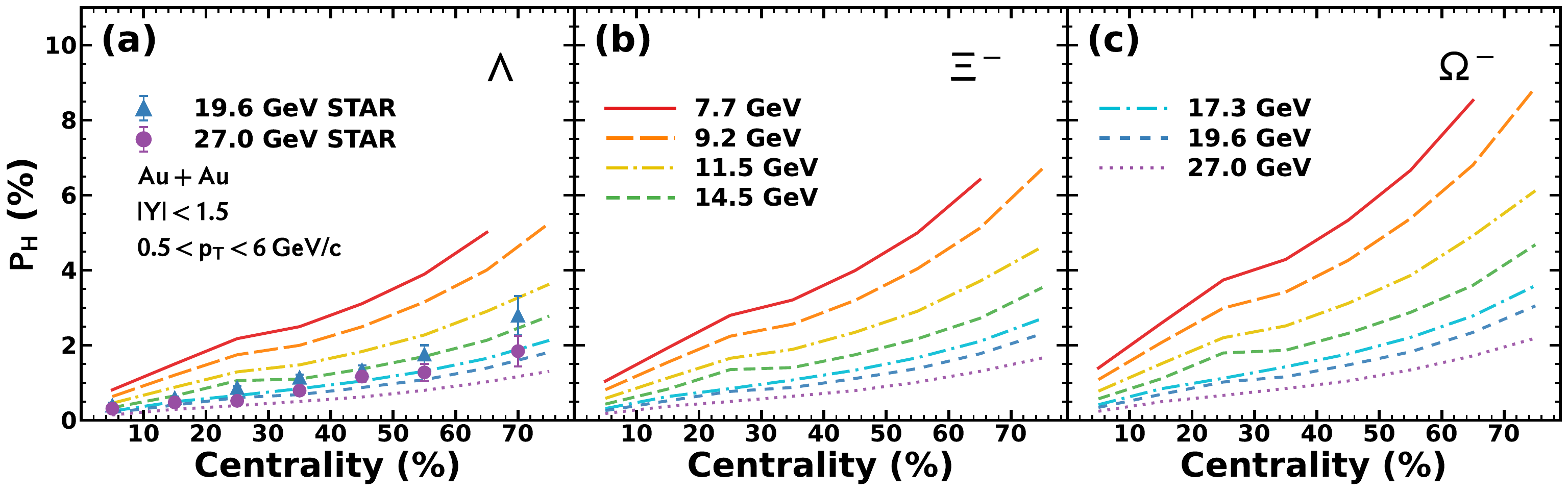} \caption{Global polarization $P_H$ of (a) $\Lambda$, (b) $\Xi^-$, and (c) $\Omega^-$ hyperons in  Au+Au collisions as a function of centrality at different collision energies. The rapidity and transverse-momentum ranges are set to $|Y|<1.5$ and $0.5<p_T<6$ GeV/c, respectively. The blue filled triangles and purple filled circles represent the experimental data points for the global polarization of $\Lambda$ at $19.6$ and $27.0$ GeV extracted from Ref.~\cite{STAR:2023nvo}, respectively. Lines with different colors and styles represent the theoretical results at different collision energies. }

\label{fig:centrality} 
\end{figure*}

To compare with experimental data, we further need to boost the polarization vector $\mathcal{P}^{\mu}$
to the rest frame of the particle,
\begin{eqnarray}
\vec{\mathcal{P}}_{*}(\mathbf{p}) & = & \vec{\mathcal{P}}(\mathbf{p})-\frac{\vec{\mathcal{P}}(\mathbf{p})\cdot\vec{p}}{p^{0}(p^{0}+m)}\vec{p}.
\label{eq:rest-frame-polarization}
\end{eqnarray}
The polarization defined above represents the primary contribution.
We still need to include the feed-down contributions discussed in Sec.~\ref{sec:feed-down}.

\section{Numerical setups}
\label{sec:numerical}

In this section, we introduce the numerical setup used in this work. We implement the (3+1)-dimensional hydrodynamic framework CLVisc with SMASH initial conditions to simulate the global spin polarization of $\Lambda$, $\Xi^-$, and $\Omega^-$ hyperons in $20$--$50\%$ Au+Au collisions at $\sqrt{s_{NN}}=7.7$--$27$ GeV.

Following our previous works~\cite{Wu:2022mkr,Yi:2023tgg}, the initial collision geometry and energy deposition are described by the hadronic transport model Simulating Many Accelerated Strongly-interacting Hadrons (SMASH) \citep{Weil:2016zrk}. 
%We collect all particles at the hydrodynamic initial time
%\begin{eqnarray}
%\tau_{0} & = & %2R/\sqrt{(\sqrt{s_{NN}}/2m_{N})^{2}-1},
%\end{eqnarray}
%which corresponds to the average nuclear overlap time.
We collect all particles at the hydrodynamic initial proper time, which is approximated by the average nuclear overlap time as
\begin{eqnarray}
\tau_{0} & = & 2R/\sqrt{(\sqrt{s_{NN}}/2m_{N})^{2}-1}.
\end{eqnarray}
Here, $R=6.38$ fm is the average radius of the Au nucleus in the Woods--Saxon distribution, $\sqrt{s_{NN}}$ is the collision energy, and $m_{N}$ is the nucleon mass. We then apply Gaussian smearing with a normalization factor $K$ and smearing widths $\sigma_{r}$ and $\sigma_{\eta}$ in the transverse and longitudinal directions, respectively~\cite{Wu:2022mkr,Yi:2023tgg}, to generate the initial energy-momentum tensor $T_{0}^{\mu\nu}$ and the net-baryon current $J_{0}^{\mu}$. Their subsequent evolution is described by the (3+1)-dimensional hydrodynamic framework CLVisc through the coupled energy-momentum and particle conservation equations. 
%\begin{eqnarray}
%\partial_{\mu}T^{\mu\nu} & = & 0,\nonumber \\
%\partial_{\mu}J^{\mu} & = & 0.
%\end{eqnarray}

\begin{table}
\centering %
\caption{Parameters used in the hydrodynamic simulations at different collision energies.}
\begin{tabular}{c|ccccc}
%\hline 
% & \multicolumn{5}{c}{SMASH IC}\tabularnewline
%\hline 
\hline 
$\sqrt{s_{NN}}$ {[}GeV{]}  & K  & $\tau_{0}$ {[}fm{]}  & $\sigma_{r}$ {[}fm{]}  & $\;\;\;\;\sigma_{\eta_{s}}\;\;$  & $\;\;C_{\eta_{v}}\;\;$\tabularnewline
\hline 
7.7  & 1.0  & 3.2  & 1.0  & 0.35  & 0.20\tabularnewline
\hline 
9.2  & 1.0  & 2.7  & 1.0  & 0.35  & 0.20\tabularnewline
\hline 
11.5  & 1.0  & 2.1  & 1.0  & 0.35  & 0.20\tabularnewline
\hline 
14.5  & 1.0  & 1.7  & 1.0  & 0.35  & 0.20\tabularnewline
\hline 
17.3  & 1.0  & 1.4  & 1.0  & 0.35  & 0.15\tabularnewline
\hline 
19.6  & 1.0  & 1.2  & 1.0  & 0.35  & 0.15\tabularnewline
\hline 
27  & 1.0  & 1.0  & 1.0  & 0.35  & 0.12\tabularnewline
\hline 
\end{tabular}
\label{tab:para} 
\end{table}

In this work, we neglect bulk viscosity and set the baryon-diffusion coefficient to $C_{B}=0$ for simplicity. We treat the specific shear viscosity $C_{\eta_{v}}$ as a free parameter. These parameters are tuned to reproduce soft-particle yields and transverse-momentum spectra at $\sqrt{s_{NN}}=7.7$, $19.6$, and $27$ GeV. We find that a single set of Gaussian smearing parameters, $K$, $\sigma_{r}$, and $\sigma_{\eta}$, is sufficient to describe the spectra at all three collision energies, suggesting that the hadronic transport model provides a reasonable description of the initial energy deposition. This is consistent with the parameters used with the Ultra-Relativistic Quantum Molecular Dynamics (UrQMD) initial conditions in Ref.~\cite{Karpenko:2015xea}. We therefore adopt the common Gaussian smearing parameters in our simulations, while choosing the same specific shear viscosity $C_{\eta_{v}}$ as that used in Ref.~\cite{Karpenko:2015xea}. All parameters used in the simulations are listed in Table~\ref{tab:para}. Further details about the model and parameter choices can be found in Refs.~\cite{Pang:2012he,Weil:2016zrk,Pang:2018zzo,Wu:2021fjf,Wu:2022mkr,Yi:2023tgg}.

\begin{figure*}[t]
\centering \includegraphics[width=0.95\linewidth]{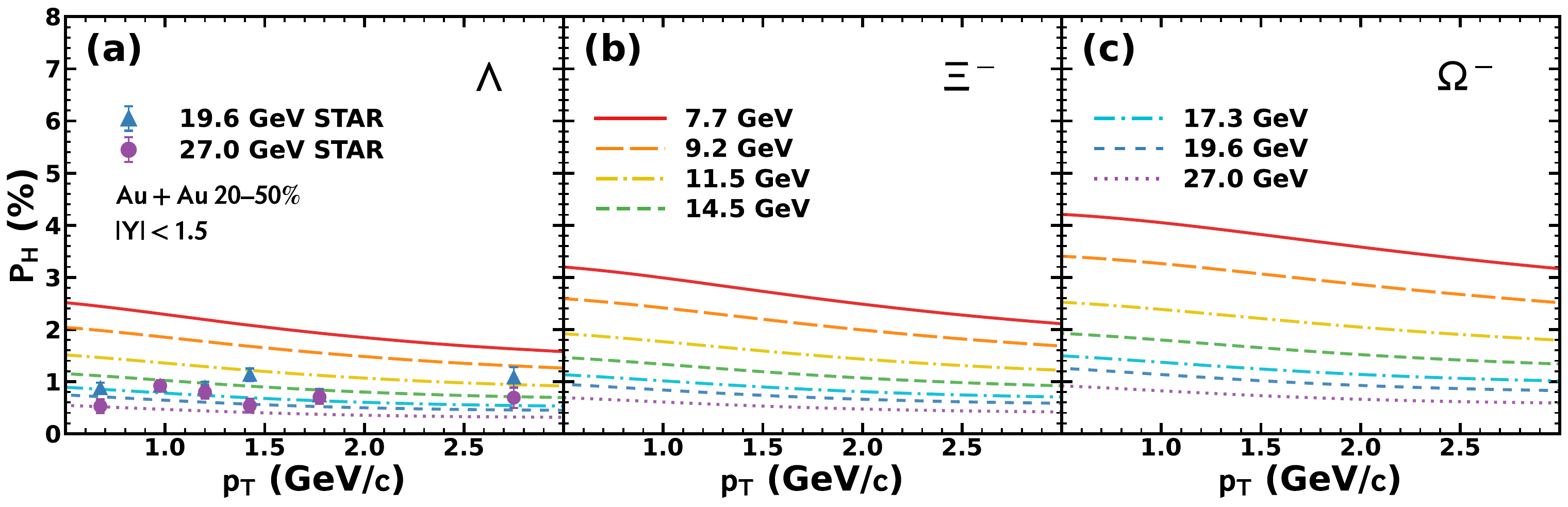}\caption{Global polarization $P_H$ of (a) $\Lambda$, (b) $\Xi^-$, and (c) $\Omega^-$ hyperons in $20$--$50\%$ central Au+Au collisions as a function of transverse momentum $p_T$ at different collision energies. The rapidity is set to $|Y|<1.5$. The blue filled triangles and purple filled circles represent the experimental data points for the global polarization of $\Lambda$ at $19.6$ and $27.0$ GeV extracted from Ref.~\cite{STAR:2023nvo}, respectively. Lines with different colors and styles represent the theoretical results at different collision energies.}
\label{fig:pT} 
\end{figure*}

To close the conservation equations, we use the NEOS-BQS equation of state \cite{Monnai:2019hkn,Monnai:2021kgu}, which smoothly connects a hadron-resonance gas at finite chemical potential to lattice-QCD results through a Taylor-expansion method. The freeze-out energy density is chosen as $\epsilon_{f}=0.4\;\text{GeV/fm}^{3}$, corresponding to $T_{f}=158$ MeV in the zero-chemical-potential limit. The relation between centrality and impact parameter $b$ is determined using the Monte Carlo Glauber model. In each centrality bin, we average over $5000$ events to obtain a smooth initial condition. After the hydrodynamic evolution, we compute the global spin polarization using the same kinematic cuts as in the experimental measurements~\cite{STAR:2023nvo,Lu:2026ghe}. The hyperon masses are $m_{\Lambda}=1.116$ GeV, $m_{\Xi}=1.322$ GeV, and $m_{\Omega}=1.672$ GeV, and the corresponding spin quantum numbers are $S=1/2$, $1/2$, and $3/2$, respectively. Therefore, within the same hydrodynamic background and polarization formula, the differences among the polarizations of these three hyperons mainly arise from their different masses and spin quantum numbers.

\section{centrality, $p_T$, and $Y$ dependent feed-down corrections \label{sec:feed-down}}

The observed $\Lambda$ and $\Xi^{-}$ samples can receive substantial contributions from decays of heavier hyperons, including $\Sigma^{0}$, $\Sigma(1385)$, and $\Xi(1530)$. These feed-down contributions can modify the measured polarizations of $\Lambda$ and $\Xi^{-}$. In experimental measurements, feed-down effects are included in the observed hyperon polarization.
Although feed-down effects on hyperon polarization have already been studied in
Refs.~\cite{Becattini:2016gvu,Karpenko:2016jyx,Li:2017slc,Xia:2019fjf,Becattini:2019ntv,Li:2021zwq},
their impact on the centrality, $p_T$, and $Y$ dependence of global polarization
in the RHIC BES-II energy range requires further investigation.
 In this section, we specify the decay kinematics, polarization transfer, and yield weighting used in our calculation.

\begin{table}[t]
\centering
\scriptsize
\setlength{\tabcolsep}{12pt}
\renewcommand{\arraystretch}{1.4}
\caption{Linear-order polarization-transfer formulas in the directly reached daughter rest frame, consistent with Refs.~\cite{Becattini:2019ntv,Xia:2019fjf}. Equation~(\ref{eq:wigner-transfer}) gives the Wigner rotation to the laboratory-defined daughter rest frame.}
\label{tab:polarization-transfer}
\begin{tabular}{@{}lc@{}}
\toprule
Spin-parity class
& $\boldsymbol{\mathcal P}_{D,\mathrm{dir}}$
\\
\midrule
$\frac12^{+}\to\frac12^{+}+1^{-}$ (EM)
& $-(\hat{\mathbf n}^{*}\!\cdot\boldsymbol{\mathcal P}_{P,*})\hat{\mathbf n}^{*}$
\\
\midrule
$\frac32^{+}\to\frac12^{+}+0^{-}$ (strong)
& $\dfrac65\boldsymbol{\mathcal P}_{P,*}-\dfrac35(\hat{\mathbf n}^{*}\!\cdot\boldsymbol{\mathcal P}_{P,*})\hat{\mathbf n}^{*}$
\\
\midrule
$\frac12^{-}\to\frac12^{+}+0^{-}$ (strong)
& $\boldsymbol{\mathcal P}_{P,*}$
\\
\midrule
$\frac12^{+}\to\frac12^{+}+0^{-}$ (weak)
& $\begin{gathered}
\left[\alpha_i+(1-\alpha_i^2)(\hat{\mathbf n}^{*}\!\cdot\boldsymbol{\mathcal P}_{P,*})\right]\hat{\mathbf n}^{*}\\[-1mm]
{}+\beta_i\boldsymbol{\mathcal P}_{P,*}\!\times\hat{\mathbf n}^{*}
+\gamma_i\hat{\mathbf n}^{*}\!\times(\boldsymbol{\mathcal P}_{P,*}\!\times\hat{\mathbf n}^{*})
\end{gathered}$
\\
\bottomrule
\end{tabular}
\end{table}

Consider a two-body decay $P\to D+X$. We denote the parent four-momentum in the laboratory frame by $P^{\mu}=(P^{0},\mathbf{P})$, with $P^{2}=M^{2}$, and the daughter four-momentum in the parent rest frame by
\begin{equation}
p^{*\mu}=m_D\left(\cosh\xi,\sinh\xi\,\hat{\mathbf n}^{*}\right),
\qquad
\hat{\mathbf n}^{*}=\frac{\mathbf p^{*}}{|\mathbf p^{*}|}.
\end{equation}
A superscript asterisk denotes a quantity evaluated in the parent rest frame. Let $\mathcal{B}(q)$ be the rotation-free boost to the rest frame of a particle with four-momentum $q^{\mu}=(q^{0},\mathbf{q})$ and mass $m_q=\sqrt{q^{2}}$:
\begin{eqnarray}
    \mathcal{B}(q)^\mu{}_{\nu} & = & \begin{pmatrix}
        \frac{q^0}{m_q} & -\frac{\mathbf{q}^{{\mathrm{T}}}}{m_q}\\
        -\frac{\mathbf{q}}{m_q} & \mathbf{1}_{3\times3}+\left(\frac{q^0}{m_q}-1\right)\hat{\mathbf{q}}\hat{\mathbf{q}}^{{\mathrm{T}}}
    \end{pmatrix},
\end{eqnarray}
where $\hat{\mathbf{q}}=\mathbf{q}/|\mathbf{q}|$, with the identity understood in the $\mathbf q=0$ limit. Thus, $\mathcal{B}^{-1}(P)$ maps the parent rest frame to the laboratory frame, and the daughter momentum in the latter is
\begin{eqnarray}
    p^{\mu} & = & \mathcal{B}^{-1}(P)^\mu{}_{\nu}p^{*\nu}.
\end{eqnarray}

The helicity-formalism results of Refs.~\cite{Becattini:2019ntv, Xia:2019fjf} can be written in the following channel-independent affine form at linear order in the normalized parent polarization:
\begin{equation}
\boldsymbol{\mathcal P}_{D,\mathrm{dir}}
=\mathbf a_i(\hat{\mathbf n}^{*})
+\mathsf T_i(\hat{\mathbf n}^{*})
\boldsymbol{\mathcal P}_{P,*}
+\mathcal{O}(|\boldsymbol{\mathcal P}_{P,*}|^{2}),
\label{eq:direct-transfer}
\end{equation}
where $\mathbf a_i$ is the polarization generated in channel $i$ from an unpolarized parent, $\mathsf T_i$ is its spin-transfer tensor, and ``dir'' denotes the daughter rest frame reached directly from the parent rest frame. We have computed the channel-dependent results listed in Table~\ref{tab:polarization-transfer}.

\begin{table*}[t]
\centering
\scriptsize
\renewcommand{\arraystretch}{1.2}
\caption{Decay channels used in the feed-down calculation. Spin-parity classes refer to Table~\ref{tab:polarization-transfer}, and EM denotes an electromagnetic decay. Inclusive branching fractions are from Ref.~\cite{ParticleDataGroup:2026aaa}; effective weights include the isospin factors described below.}
\label{tab:feeddown-channels}
\begin{tabular*}{0.94\textwidth}{@{\extracolsep{\fill}}cccccc@{}}
\toprule
Parent
& Decay channel
& Decay type
& Spin-parity class
& Effective branching weight
& $w_i(\Omega^{*})$
\\
\midrule
$\Sigma^{0}$
& $\Lambda+\gamma$
& EM
& $\frac12^{+}\to\frac12^{+}+1^{-}$
& $100\%$
& $1/(4\pi)$
\\
\midrule
\multirow{2}{*}{$\Sigma(1385)$}
& $\Lambda+\pi$
& \multirow{2}{*}{Strong}
& \multirow{2}{*}{$\frac32^{+}\to\frac12^{+}+0^{-}$}
& $87\%$
& $1/(4\pi)$
\\
& $\Sigma^{0}+\pi$
&
&
& $11.7\%\times\frac13$
& $1/(4\pi)$
\\
\midrule
$\Lambda(1405)$
& $\Sigma^{0}+\pi^{0}$
& Strong
& $\frac12^{-}\to\frac12^{+}+0^{-}$
& $100\%\times\frac13$
& $1/(4\pi)$
\\
\midrule
$\Xi^{Q}$ ($Q=0,-1$)
& $\Lambda+\pi^{Q}$
& Weak
& $\frac12^{+}\to\frac12^{+}+0^{-}$
& $100\%$
& $\dfrac{1+\alpha_{\Xi^{Q}}\boldsymbol{\mathcal P}_{\Xi^{Q},*}\cdot\hat{\mathbf n}^{*}}{4\pi}$
\\
\midrule
$\Xi(1530)$
& $\Xi+\pi$
& Strong
& $\frac32^{+}\to\frac12^{+}+0^{-}$
& $100\%$
& $1/(4\pi)$
\\
\bottomrule
\end{tabular*}
\end{table*}
%\FloatBarrier

The directly reached daughter rest frame and the daughter rest frame reached through the laboratory frame generally differ by a Wigner rotation. Therefore, the polarization entering the laboratory-defined observable is
\begin{align}
\boldsymbol{\mathcal P}_{D,\mathrm{drf}}
&=\mathsf R_W(P,p^{*})
\boldsymbol{\mathcal P}_{D,\mathrm{dir}},\nonumber\\
\mathcal R_W(P,p^{*})
&=\mathcal B(p)\mathcal B^{-1}(P)\mathcal B^{-1}(p^{*}),
\label{eq:wigner-transfer}
\end{align}
where $\mathsf R_W$ is the spatial part of the little-group transformation $\mathcal R_W$. Thus, $\mathcal P_{D,\mathrm{drf}}^{y}$ is the feed-down counterpart of the rest-frame component $\mathcal P_{*}^{y}$ in Eq.~(\ref{eq:rest-frame-polarization}).

%For a weak channel, the Lee--Yang parameters satisfy $\alpha_i^2+\beta_i^2+\gamma_i^2=1$. {\color{blue}[SP: What is the definition of these angles?]} 
For a weak channel, the Lee--Yang parameters $\alpha_i$, $\beta_i$, and $\gamma_i$
characterize the decay asymmetry and polarization transfer~\cite{Lee:1957qs}.
The parameter $\alpha_i$ governs the asymmetry of the daughter angular distribution
relative to the parent polarization direction  and contributes to the parent-polarization-independent component of the daughter polarization.
The parameters $\beta_i$ and $\gamma_i$ govern the transfer of polarization to the daughter
in the directions normal to and within the plane spanned by the parent polarization
and daughter momentum in the parent rest frame, respectively,
with both directions perpendicular to the daughter momentum.
These parameters satisfy $\alpha_i^2+\beta_i^2+\gamma_i^2=1$.For $\Xi^Q\to\Lambda\pi^Q$ ($Q=0,-1$) channel, we use the decay parameters
$(\alpha_{\Xi^Q},\beta_{\Xi^Q},\gamma_{\Xi^Q})$ obtained from Particle Data Group (PDG)~\cite{ParticleDataGroup:2026aaa}.
% For $\Xi^{Q}\to\Lambda\pi^{Q}$, we use the charge-specific set $(\alpha_{\Xi^Q},\beta_{\Xi^Q},\gamma_{\Xi^Q})$, with $Q=0,-1$.

For each decay channel $i$, we define the normalized daughter angular distribution in the parent rest frame by
\begin{eqnarray}
    w_i(\Omega^{*}) & \equiv & \frac{1}{N_i}\frac{dN_i}{d\Omega^{*}},
    \qquad
    \int d\Omega^{*}\,w_i(\Omega^{*})=1,
\end{eqnarray}
where $N_i$ is the total number of decays, $\Omega^{*}=(\theta^{*},\phi^{*})$, and $d\Omega^{*}=d\cos\theta^{*}\,d\phi^{*}$. At the order retained here, the strong and electromagnetic (EM) channels in Table~\ref{tab:feeddown-channels} are isotropic when higher-rank spin alignment is neglected, so that $w_i(\Omega^{*})=1/(4\pi)$. For the weak decay $\Xi^{Q}\to\Lambda\pi^{Q}$,
\begin{equation}
w_i(\Omega^{*})
=\frac{1+\alpha_{\Xi^{Q}}\boldsymbol{\mathcal P}_{\Xi^{Q},*}\cdot\hat{\mathbf n}^{*}}{4\pi}.
\end{equation}

The entries in Table~\ref{tab:feeddown-channels} are elementary decay steps. For a cascade, the angular folding and polarization transfer are applied recursively, with the daughter spectrum and polarization from one step serving as the parent input to the next.

We neglect the small mass and primary-yield differences among charge states within an isospin multiplet.
An individual charge state has degeneracy $g_a=2J_a+1$, with $J_a$ the spin quantum number of the particle $a$.
After equally weighted charge states are combined, the multiplet degeneracy is $g_a=(2J_a+1)(2I_a+1)$, with $I_a$ is the isospin quantum number of the multiplet.
The weak-decay parameters are also averaged over the charge states in the calculation.

%We neglect the small mass and primary-yield differences among charge states within an isospin multiplet. An individual charge state has degeneracy $g_a=2J_a+1$ with $J_a$ the spin of the particle $a$; after equally weighted charge states are combined, the multiplet degeneracy is $g_a=(2J_a+1)(2I_a+1)$. The weak-decay parameters are also averaged over the charge states in the calculation.

% Under isospin conservation, the Clebsch--Gordan coefficients give probabilities $(1/2,0,1/2)$ for the $(+,0,-)$ charge states of $\Sigma(1385)$, respectively, to decay into a final state containing $\Sigma^{0}$. 
Under isospin conservation, the Clebsch--Gordan coefficients
give fractions $(1/2,0,1/2)$ of $\Sigma(1385)\to\Sigma\pi$
decays producing $\Sigma^0$ for the parent charge states
$(+,0,-)$, respectively. Equal weighting of the three parent charge states gives an average factor $1/3$ and hence the effective weight $11.7\%\times(1/3)$ in Table~\ref{tab:feeddown-channels}. For $\Lambda(1405)\to\Sigma^{0}\pi^{0}$, the factor $1/3$ follows from the three equal $\Sigma\pi$ charge channels of the isosinglet decay

The final global polarization is obtained by combining the primary contribution of the observed species with all of its feed-down contributions. For a directly emitted species $a$ with four-momentum $q^{\mu}$, we denote the differential yield by $\mathcal{Y}_a(q)$ and evaluate it with the Cooper--Frye formula,
\begin{eqnarray}
    \mathcal{Y}_a(q) & \equiv & \frac{dN_a}{dY\,q_T\,dq_T\,d\phi_{q}}\nonumber\\
     & = & \frac{g_a}{(2\pi)^3}\int d\Sigma\cdot q\,f_{{\rm eq},a}(q,x),
\end{eqnarray}
where $g_a$ and $f_{{\rm eq},a}$ are the degeneracy factor and equilibrium distribution function of species $a$, respectively. We write the daughter yield from decay step $i$ as
\begin{eqnarray}
    \mathcal{Y}^{\mathrm{fd}}_{i}(p)
    & \equiv &
    \mathrm{Br}_i\,
    \mathcal{F}_i\!\left[\mathcal{Y}_{P_i}\right](p),
\end{eqnarray}
where $\mathcal{Y}_{P_i}$ is the input spectrum of parent species $P_i$, either primary or generated by a preceding decay, $\mathrm{Br}_i$ is the effective branching weight, and $\mathcal{F}_i$ folds the parent spectrum over $\Omega^{*}$ with $w_i(\Omega^{*})$ and the decay kinematics. At fixed daughter momentum $p$, the channel polarization four-vector is the corresponding yield-weighted conditional average,
\begin{equation}
\mathcal{P}^{\mu}_{i}(p)
\equiv
\frac{\mathcal{F}_i\!\left[\mathcal{Y}_{P_i}\mathcal{P}^{\mu}_{D,\mathrm{drf},i}\right](p)}
{\mathcal{F}_i\!\left[\mathcal{Y}_{P_i}\right](p)}.
\label{eq:conditional-feeddown-polarization}
\end{equation}
We then define the polarization-weighted four-vector spectrum $\mathcal{N}^{\mu}(p)$ and the scalar total spectrum $\mathcal{D}(p)$ as
\begin{align}
\mathcal{N}^{\mu}(p)
={}&
\mathcal{Y}^{\mathrm{prim}}(p)
\mathcal{P}^{\mu}_{\mathrm{prim}}(p)
+
\sum_i
\mathcal{Y}^{\mathrm{fd}}_{i}(p)
\mathcal{P}^{\mu}_{i}(p),\\
\mathcal{D}(p)
={}&
\mathcal{Y}^{\mathrm{prim}}(p)
+\sum_i
\mathcal{Y}^{\mathrm{fd}}_{i}(p).
\end{align}
Here, $\mathcal{Y}^{\mathrm{prim}}(p)$ is the primary yield, and $\mathcal{P}^{\mu}_{\mathrm{prim}}(p)\equiv\mathcal{P}^{\mu}_{*}(p)$ is the primary rest-frame polarization four-vector. It is obtained by first evaluating Eq.~(\ref{eq:modifed-CF}) and then applying the rest-frame transformation in Eq.~(\ref{eq:rest-frame-polarization}), with $\mathcal{P}^{\mu}_{*}=(0,\vec{\mathcal{P}}_{*})$. The species dependence is implicit in the primary term and in the set of channels $i$; any preceding cascade steps are already included recursively in $\mathcal{Y}_{P_i}$ and $\mathcal{P}^{\mu}_{i}$.

The feed-down-corrected integrated, $p_T$-differential, and $Y$-differential polarizations are
\begin{eqnarray}
P_{H}
& = &-
\frac{\displaystyle\int p_T\,dp_T\,dY\,d\phi_p\,\mathcal{N}^{y}(p)}
{\displaystyle\int p_T\,dp_T\,dY\,d\phi_p\,\mathcal{D}(p)},\\
P_{H}(p_T)
& = &-
\frac{\displaystyle\int dY\,d\phi_p\,\mathcal{N}^{y}(p)}
{\displaystyle\int dY\,d\phi_p\,\mathcal{D}(p)},\\
P_{H}(Y)
& = &-
\frac{\displaystyle\int p_T\,dp_T\,d\phi_p\,\mathcal{N}^{y}(p)}
{\displaystyle\int p_T\,dp_T\,d\phi_p\,\mathcal{D}(p)}.
\end{eqnarray}
For convenience, we also define the global-polarization
splitting between hyperon \mbox{$H$} and its antiparticle
\mbox{$\bar{H}$} as
\begin{equation}
\Delta P_H \equiv P_{\overline{H}} - P_H.
\end{equation}

\section{Results and discussion}
\label{sec:Results-and-discussion}

We now discuss our results for the global polarization of $\Lambda$, $\Xi^{-}$, and $\Omega^{-}$ hyperons as functions of collision energy, centrality, transverse momentum, and rapidity. 

In Fig.~\ref{fig:energy}(a), we show the global polarization $P_H$ as functions of collision energy for $\Lambda$, $\Xi^-$, and $\Omega^-$ hyperons 
%\comment{$\Omega$? and $\Xi$? because both $\Omega^+$ and  $\Omega^-$ are included in fig 1, if so, the caption of fig 1 also needs to be fixed. }
We first observe that the global polarization of all three hyperon species increases as the collision energy decreases. This trend can be understood as a consequence of stronger net-baryon stopping, which deposits a larger fraction of the initial angular momentum into the produced medium, together with the smaller particle multiplicity near midrapidity at lower collision energies.

\begin{figure*}[t]
\centering \includegraphics[width=0.95\linewidth]{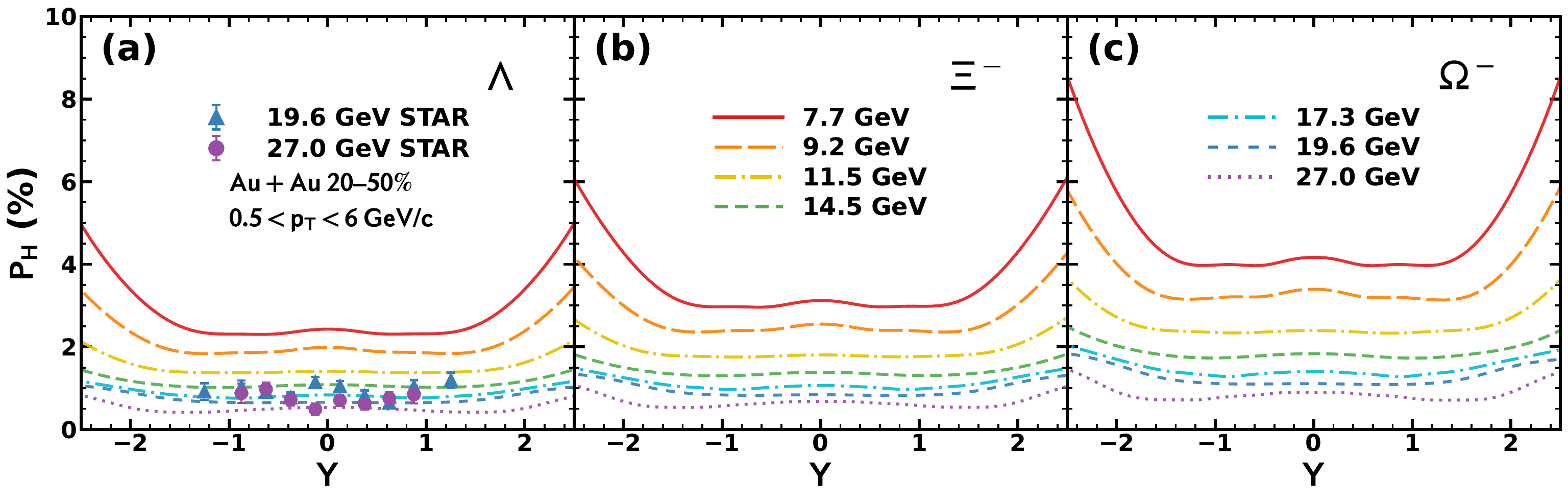}\caption{Global polarization $P_H$ of (a) $\Lambda$, (b) $\Xi^-$, and (c) $\Omega^-$ hyperons in $20$--$50\%$ central Au+Au collisions as a function of rapidity $Y$ at different collision energies. The transverse-momentum ranges are set to $0.5<p_T<6$ GeV/c. The blue filled triangles and purple filled circles represent the experimental data points for the global polarization of $\Lambda$ at $19.6$ and $27.0$ GeV extracted from Ref.~\cite{STAR:2023nvo}, respectively. Lines with different colors and styles represent the theoretical results at different collision energies.}
\label{fig:rapidity} 
\end{figure*}

We next examine the global polarization of each hyperon species. For $\Lambda$ and $\Xi^-$ %\comment{$\Xi$}
hyperons, since they have similar masses and the same spin quantum number, their primary polarizations computed from Eq.~(\ref{eq:modifed-CF}) are close to each other.
On the other hand, according to Ref. ~\cite{Becattini:2016gvu,Karpenko:2016jyx,Li:2017slc,Li:2021zwq}, feed-down effects enhance the $\Xi^-$ polarization while reducing the $\Lambda$ polarization. After including feed-down effects, our results for the global polarization of $\Xi^-$ hyperons are consistent with the experimental data, while those for $\Lambda$ hyperons are close to the data but slightly smaller at intermediate collision energies.
We now discuss the global polarization of the $\Omega^-$ hyperon. Since the spin of the $\Omega^-$ hyperon is $3/2$, Eq.~(\ref{eq:modifed-CF}) gives the simple estimate $P_\Omega \approx 5P_\Lambda/3$ if the mass difference between $\Omega^-$ and $\Lambda$ and feed-down effects are neglected \cite{Liang:2004ph,Lv:2024uev, Zhang:2024hyq,Lv:2026kct}. Our numerical results are consistent with this estimate and show an enhancement of the global polarization of $\Omega^-$. Nevertheless, our results remain well below the experimental central value, similar to earlier studies based on kinetic approaches \cite{Li:2021zwq}. This discrepancy may indicate that additional mechanisms, such as two- and three-particle spin-spin correlations among strange quarks inside the $\Omega^-$ hyperon, could contribute to the observed $\Omega^-$ polarization \cite{Lv:2024uev, Zhang:2024hyq,Lv:2026kct}.

In Fig.~\ref{fig:energy}(b), we present the global-polarization splitting $\Delta P_H$ as a function of collision energy. We find that anti-hyperons systematically carry a larger global polarization than the corresponding hyperons. This difference originates mainly from the chemical-potential-gradient contribution in Eq.~(\ref{eq:modifed-CF}). Therefore, a possible splitting of the global polarization between hyperons and anti-hyperons can arise not only from electromagnetic fields but also from the spin Hall effect. This finding is consistent with previous studies \cite{Ryu:2021lnx}. Our results for $\Delta P_\Lambda$ are consistent with the experimental data within uncertainties. We also provide predictions for $\Delta P_\Xi$ and $\Delta P_\Omega$.

Next, we present the global polarization of the three hyperon species as functions of centrality, transverse momentum $p_{T}$, and rapidity $Y$ in Figs.~\ref{fig:centrality}, \ref{fig:pT}, and \ref{fig:rapidity}, respectively. %Due to the lack of centrality-, $p_T$-, and $Y$-dependent feed-down corrections to the global polarization, we neglect feed-down effects for the remaining differential observables. 

In Fig.~\ref{fig:centrality}, we observe that the centrality dependence of the global polarization of hyperons at intermediate collision energies is similar to that found at high collision energies \cite{STAR:2018gyt,STAR:2020xbm,ALICE:2019onw,STAR:2025dgs}. Namely, the polarization increases toward more peripheral collisions. This behavior can be attributed to the larger vorticity generated by stronger velocity gradients in more peripheral collisions, which enhances the global polarization. We also find that, at a fixed centrality, the global polarization becomes larger as the collision energy decreases, consistent with the collision-energy dependence shown in Fig.~\ref{fig:energy}.

In Fig.~\ref{fig:pT}, we observe that the global polarization decreases with increasing $p_{T}$. This trend is consistent with that observed in the experimental data. In Fig.~\ref{fig:rapidity}, we also find that the global polarization of hyperons is nearly flat around midrapidity $|Y|\in[0,1]$ and then increases rapidly with increasing $|Y|$ for $|Y|\geq1$.

Comparing the global polarization of different hyperons in Figs.~\ref{fig:centrality}, \ref{fig:pT}, and \ref{fig:rapidity}, we observe that $P_\Xi \gtrsim P_\Lambda$ and $P_\Omega \simeq 5P_\Lambda/3$ at fixed collision energy, centrality, $p_T$, and $Y$, consistent with the spin and mass dependence in Eq.~(\ref{eq:modifed-CF}).

\section{Summary}
\label{sec:Summary}

In this work, we have studied the global spin polarization of $\Lambda$, $\Xi^{-}$, and $\Omega^{-}$
hyperons, including feed-down effects, in $20$--$50\%$ Au+Au collisions over $\sqrt{s_{NN}}=7.7$--$27$ GeV,
using the modified Cooper--Frye formula in Eq.~(\ref{eq:modifed-CF}) combined with the
(3+1)-dimensional viscous hydrodynamic framework CLVisc with SMASH initial conditions.

We first show the global polarization of hyperons, $P_H$, as a function of collision energy in Fig.~\ref{fig:energy}(a). We find that the global polarization of all three hyperon species increases as the collision energy decreases, in agreement with the trend observed by STAR.
We observe a slight difference between the global polarizations of $\Xi^{-}$ and $\Lambda$ due to feed-down effects. In particular, feed-down effects enhance the global polarization of $\Xi^{-}$ but reduce that of $\Lambda$.
The global polarization of $\Omega^{-}$ is systematically larger than those of $\Lambda$ and $\Xi^{-}$ owing to its larger spin quantum number. Nevertheless, our result for the global polarization of $\Omega^{-}$ still underestimates the central value of the recent STAR measurement. This discrepancy may indicate the presence of two- and three-particle short-range spin correlations among the strange quarks inside the $\Omega^{-}$.

In Fig.~\ref{fig:energy}(b), we also present the global-polarization splitting $\Delta P_H$ as a function of collision energy. We find that the global polarization of anti-hyperons is larger than that of the corresponding hyperons. The splitting increases toward lower collision energies and is dominated by the chemical-potential-gradient contribution. Thus, the spin Hall effect can influence the global polarization in a way qualitatively similar to the electromagnetic-field contribution.

Finally, we plot the global polarization of hyperons as functions of centrality, transverse momentum $p_{T}$, and rapidity $Y$ in Figs.~\ref{fig:centrality}, \ref{fig:pT}, and \ref{fig:rapidity}, respectively. For all three hyperon species, the global polarization increases toward more peripheral collisions, decreases with increasing transverse momentum, and is nearly flat around midrapidity.

Our results provide a systematic theoretical baseline for the global polarization of multi-strange hyperons across the RHIC BES-II energy range. They also suggest that short-range spin correlations among strange quarks inside the $\Omega^{-}$ may be a promising direction for future theoretical and experimental investigation.

\section*{Acknowledgment}
We would like to thank Zhenyu Chen, Xingrui Gou and Tong Fu  for helpful discussion.
This work is supported in part by the National Key Research and Development
Program of China under Contract No. 2022YFA1605500, by the Chinese
Academy of Sciences (CAS) under Grants No. YSBR-088 and by National Natural Science Foundation of China (NSFC) under Grant Nos. 12075235,
12225503, 11890710, 11890711, 11935007, 12175122, 2021-867, 11221504,
11861131009,12547164 and 11890714,
by the China Postdoctoral Science Foundation under Grant Number 2025M783388.
X.-Y.W. is supported in part by the Natural Sciences and Engineering Research Council of Canada
(NSERC) [SAPIN-2026-00047and SAPIN-2024-00026].

\bibliographystyle{h-physrev}
\bibliography{qkt-ref-2}

@article{STAR:2023nvo,
    author = "Abdulhamid, M. I. and others",
    collaboration = "STAR",
    title = "{Global polarization of \ensuremath{\Lambda} and \ensuremath{\Lambda}\textasciimacron{} hyperons in Au+Au collisions at $\sqrt{s_\mathrm{NN}}$=19.6 and 27 GeV}",
    eprint = "2305.08705",
    archivePrefix = "arXiv",
    primaryClass = "nucl-ex",
    doi = "10.1103/PhysRevC.108.014910",
    journal = "Phys. Rev. C",
    volume = "108",
    number = "1",
    pages = "014910",
    year = "2023"
}

@article{ALICE:2019onw,
    author = "Acharya, Shreyasi and others",
    collaboration = "ALICE",
    title = "{Global polarization of $\Lambda \bar \Lambda$ hyperons in Pb-Pb collisions at $\sqrt {s_{NN}}$ = 2.76 and 5.02 TeV}",
    eprint = "1909.01281",
    archivePrefix = "arXiv",
    primaryClass = "nucl-ex",
    reportNumber = "CERN-EP-2019-173",
    doi = "10.1103/PhysRevC.101.044611",
    journal = "Phys. Rev. C",
    volume = "101",
    number = "4",
    pages = "044611",
    year = "2020",
    note = "[Erratum: Phys.Rev.C 105, 029902 (2022)]"
}

@article{Xu:2022hql,
    author = "Xu, Kun and Lin, Fan and Huang, Anping and Huang, Mei",
    title = "{\ensuremath{\Lambda}/\ensuremath{\Lambda}\textasciimacron{} polarization and splitting induced by rotation and magnetic field}",
    eprint = "2205.02420",
    archivePrefix = "arXiv",
    primaryClass = "hep-ph",
    doi = "10.1103/PhysRevD.106.L071502",
    journal = "Phys. Rev. D",
    volume = "106",
    number = "7",
    pages = "L071502",
    year = "2022"
}

@article{Buzzegoli:2022qrr,
    author = "Buzzegoli, Matteo",
    title = "{Spin polarization induced by magnetic field and the relativistic Barnett effect}",
    eprint = "2211.04549",
    archivePrefix = "arXiv",
    primaryClass = "nucl-th",
    doi = "10.1016/j.nuclphysa.2023.122674",
    journal = "Nucl. Phys. A",
    volume = "1036",
    pages = "122674",
    year = "2023"
}

@article{Buzzegoli:2022fxu,
    author = "Buzzegoli, M. and Becattini, F. and Inghirami, G. and Karpenko, I. and Palermo, A.",
    title = "{Spin-thermal Shear Coupling in Relativistic Nuclear Collisions}",
    eprint = "2208.04449",
    archivePrefix = "arXiv",
    primaryClass = "nucl-th",
    doi = "10.5506/APhysPolBSupp.16.1-A39",
    journal = "Acta Phys. Polon. Supp.",
    volume = "16",
    number = "1",
    pages = "39",
    year = "2023"
}

@article{Becattini:2022zvf,
    author = "Becattini, Francesco",
    title = "{Spin and polarization: a new direction in relativistic heavy ion physics}",
    eprint = "2204.01144",
    archivePrefix = "arXiv",
    primaryClass = "nucl-th",
    doi = "10.1088/1361-6633/ac97a9",
    journal = "Rept. Prog. Phys.",
    volume = "85",
    number = "12",
    pages = "122301",
    year = "2022"
}

@article{Wu:2022mkr,
    author = "Wu, Xiang-Yu and Yi, Cong and Qin, Guang-You and Pu, Shi",
    title = "{Local and global polarization of \ensuremath{\Lambda} hyperons across RHIC-BES energies: The roles of spin hall effect, initial condition, and baryon diffusion}",
    eprint = "2204.02218",
    archivePrefix = "arXiv",
    primaryClass = "hep-ph",
    doi = "10.1103/PhysRevC.105.064909",
    journal = "Phys. Rev. C",
    volume = "105",
    number = "6",
    pages = "064909",
    year = "2022"
}

@article{Florkowski:2021xvy,
    author = "Florkowski, Wojciech and Kumar, Avdhesh and Mazeliauskas, Aleksas and Ryblewski, Radoslaw",
    title = "{Effect of thermal shear on longitudinal spin polarization in a thermal model}",
    eprint = "2112.02799",
    archivePrefix = "arXiv",
    primaryClass = "hep-ph",
    doi = "10.1103/PhysRevC.105.064901",
    journal = "Phys. Rev. C",
    volume = "105",
    number = "6",
    pages = "064901",
    year = "2022"
}

@article{Muller:2021hpe,
    author = {M\"uller, Berndt and Yang, Di-Lun},
    title = "{Anomalous spin polarization from turbulent color fields}",
    eprint = "2110.15630",
    archivePrefix = "arXiv",
    primaryClass = "nucl-th",
    doi = "10.1103/PhysRevD.105.L011901",
    journal = "Phys. Rev. D",
    volume = "105",
    number = "1",
    pages = "L011901",
    year = "2022",
    note = "[Erratum: Phys.Rev.D 106, 039904 (2022)]"
}

@article{Becattini:2021suc,
    author = "Becattini, F. and Buzzegoli, M. and Palermo, A.",
    title = "{Spin-thermal shear coupling in a relativistic fluid}",
    eprint = "2103.10917",
    archivePrefix = "arXiv",
    primaryClass = "nucl-th",
    doi = "10.1016/j.physletb.2021.136519",
    journal = "Phys. Lett. B",
    volume = "820",
    pages = "136519",
    year = "2021"
}

@article{STAR:2018gyt,
    author = "Adam, Jaroslav and others",
    collaboration = "STAR",
    title = "{Global polarization of $\Lambda$ hyperons in Au+Au collisions at $\sqrt{s_{_{NN}}}$ = 200 GeV}",
    eprint = "1805.04400",
    archivePrefix = "arXiv",
    primaryClass = "nucl-ex",
    doi = "10.1103/PhysRevC.98.014910",
    journal = "Phys. Rev. C",
    volume = "98",
    pages = "014910",
    year = "2018"
}

@Article{STAR:2017ckg,
  Title                    = {{Global $\Lambda$ hyperon polarization in nuclear collisions: evidence for the most vortical fluid}},
  Author                   = {Adamczyk, L. and others},
  DOI                      = {10.1038/nature23004},
  Eprint                   = {1701.06657},
  Pages                    = {62-65},
  Volume                   = {548},
  Year                     = {2017},
  Archiveprefix            = {arXiv},
  Collaboration            = {STAR},
  Journal                  = {Nature},
  Primaryclass             = {nucl-ex},
  Slaccitation             = {%%CITATION = ARXIV:1701.06657;%%}
}

@article{Li:2021zwq,
    author = "Li, Hui and Xia, Xiao-Liang and Huang, Xu-Guang and Huang, Huan Zhong",
    title = "{Global spin polarization of multistrange hyperons and feed-down effect in heavy-ion collisions}",
    eprint = "2106.09443",
    archivePrefix = "arXiv",
    primaryClass = "nucl-th",
    doi = "10.1016/j.physletb.2022.136971",
    journal = "Phys. Lett. B",
    volume = "827",
    pages = "136971",
    year = "2022"
}

@article{Becattini:2019ntv,
    author = "Becattini, Francesco and Cao, Gaoqing and Speranza, Enrico",
    title = "{Polarization transfer in hyperon decays and its effect in relativistic nuclear collisions}",
    eprint = "1905.03123",
    archivePrefix = "arXiv",
    primaryClass = "nucl-th",
    doi = "10.1140/epjc/s10052-019-7213-6",
    journal = "Eur. Phys. J. C",
    volume = "79",
    number = "9",
    pages = "741",
    year = "2019"
}

@Article{Becattini:2013fla,
  Title                    = {{Relativistic distribution function for particles with spin at local thermodynamical equilibrium}},
  Author                   = {Becattini, F. and Chandra, V. and Del Zanna, L. and Grossi, E.},
  DOI                      = {10.1016/j.aop.2013.07.004},
  Eprint                   = {1303.3431},
  Pages                    = {32--49},
  Volume                   = {338},
  Year                     = {2013},
  Archiveprefix            = {arXiv},
  Journal                  = {Annals Phys.},
  Primaryclass             = {nucl-th}
}

@Article{Becattini:2013vja,
  Title                    = {{$\Lambda$ polarization in peripheral heavy ion collisions}},
  Author                   = {Becattini, F. and Csernai, L. and Wang, D.J.},
  DOI                      = {10.1103/PhysRevC.88.034905},
  Eprint                   = {1304.4427},
  Note                     = {[Erratum: Phys.Rev.C 93, 069901 (2016)]},
  Number                   = {3},
  Pages                    = {034905},
  Volume                   = {88},
  Year                     = {2013},
  Archiveprefix            = {arXiv},
  Journal                  = {Phys. Rev. C},
  Primaryclass             = {nucl-th}
}

@Article{Becattini:2007nd,
  Title                    = {{The Ideal relativistic spinning gas: Polarization and spectra}},
  Author                   = {Becattini, F. and Piccinini, F.},
  DOI                      = {10.1016/j.aop.2008.01.001},
  Eprint                   = {0710.5694},
  Pages                    = {2452--2473},
  Volume                   = {323},
  Year                     = {2008},
  Archiveprefix            = {arXiv},
  Journal                  = {Annals Phys.},
  Primaryclass             = {nucl-th}
}

@Article{Betz:2007kg,
  Title                    = {{Polarization probes of vorticity in heavy ion collisions}},
  Author                   = {Betz, Barbara and Gyulassy, Miklos and Torrieri, Giorgio},
  DOI                      = {10.1103/PhysRevC.76.044901},
  Eprint                   = {0708.0035},
  Pages                    = {044901},
  Volume                   = {76},
  Year                     = {2007},
  Archiveprefix            = {arXiv},
  Journal                  = {Phys. Rev. C},
  Primaryclass             = {nucl-th}
}

@Article{Csernai:2013bqa,
  Title                    = {{Flow Vorticity in Peripheral High Energy Heavy Ion Collisions}},
  Author                   = {Csernai, L.P. and Magas, V.K. and Wang, D.J.},
  DOI                      = {10.1103/PhysRevC.87.034906},
  Eprint                   = {1302.5310},
  Number                   = {3},
  Pages                    = {034906},
  Volume                   = {87},
  Year                     = {2013},
  Archiveprefix            = {arXiv},
  Journal                  = {Phys. Rev. C},
  Primaryclass             = {nucl-th}
}

@Article{Gao:2020vbh,
  Title                    = {{Recent developments in chiral and spin polarization effects in heavy-ion collisions}},
  Author                   = {Gao, Jian-Hua and Ma, Guo-Liang and Pu, Shi and Wang, Qun},
  DOI                      = {10.1007/s41365-020-00801-x},
  Eprint                   = {2005.10432},
  Number                   = {9},
  Pages                    = {90},
  Volume                   = {31},
  Year                     = {2020},
  Archiveprefix            = {arXiv},
  Journal                  = {Nucl. Sci. Tech.},
  Primaryclass             = {hep-ph}
}

@Article{Guo:2019joy,
  Title                    = {{Magnetic Field Induced Polarization Difference between Hyperons and Anti-hyperons}},
  Author                   = {Guo, Yu and Shi, Shuzhe and Feng, Shengqin and Liao, Jinfeng},
  DOI                      = {10.1016/j.physletb.2019.134929},
  Eprint                   = {1905.12613},
  Pages                    = {134929},
  Volume                   = {B798},
  Year                     = {2019},
  Archiveprefix            = {arXiv},
  Journal                  = {Phys. Lett.},
  Primaryclass             = {nucl-th},
  Slaccitation             = {%%CITATION = ARXIV:1905.12613;%%}
}

@Article{Hidaka:2017auj,
  Title                    = {{Nonlinear Responses of Chiral Fluids from Kinetic Theory}},
  Author                   = {Hidaka, Yoshimasa and Pu, Shi and Yang, Di-Lun},
  DOI                      = {10.1103/PhysRevD.97.016004},
  Eprint                   = {1710.00278},
  Number                   = {1},
  Pages                    = {016004},
  Volume                   = {D97},
  Year                     = {2018},
  Archiveprefix            = {arXiv},
  Journal                  = {Phys. Rev.},
  Primaryclass             = {hep-th},
  Reportnumber             = {RIKEN-QHP-260, RIKEN-iTHEMS-Report-17},
  Slaccitation             = {%%CITATION = ARXIV:1710.00278;%%}
}

@article{Ivanov:2019ern,
    author = "Ivanov, Yu B. and Toneev, V. D. and Soldatov, A. A.",
    title = "{Estimates of hyperon polarization in heavy-ion collisions at collision energies $\sqrt{s_{NN}}=$ 4--40 GeV}",
    eprint = "1903.05455",
    archivePrefix = "arXiv",
    primaryClass = "nucl-th",
    doi = "10.1103/PhysRevC.100.014908",
    journal = "Phys. Rev. C",
    volume = "100",
    number = "1",
    pages = "014908",
    year = "2019"
}

@article{Karpenko:2016jyx,
    author = "Karpenko, I. and Becattini, F.",
    title = "{Study of $\Lambda $ polarization in relativistic nuclear collisions at $\sqrt{s_\mathrm {NN}}=7.7$ \textendash{}200 GeV}",
    eprint = "1610.04717",
    archivePrefix = "arXiv",
    primaryClass = "nucl-th",
    doi = "10.1140/epjc/s10052-017-4765-1",
    journal = "Eur. Phys. J. C",
    volume = "77",
    number = "4",
    pages = "213",
    year = "2017"
}

@article{Shi:2017wpk,
    author = "Shi, Shuzhe and Li, Kangle and Liao, Jinfeng",
    title = "{Searching for the Subatomic Swirls in the CuCu and CuAu Collisions}",
    eprint = "1712.00878",
    archivePrefix = "arXiv",
    primaryClass = "nucl-th",
    doi = "10.1016/j.physletb.2018.09.066",
    journal = "Phys. Lett. B",
    volume = "788",
    pages = "409--413",
    year = "2019"
}

@article{Liu:2019krs,
    author = "Liu, Shuai Y. F. and Sun, Yifeng and Ko, Che Ming",
    title = "{Spin Polarizations in a Covariant Angular-Momentum-Conserved Chiral Transport Model}",
    eprint = "1910.06774",
    archivePrefix = "arXiv",
    primaryClass = "nucl-th",
    doi = "10.1103/PhysRevLett.125.062301",
    journal = "Phys. Rev. Lett.",
    volume = "125",
    number = "6",
    pages = "062301",
    year = "2020"
}

@article{Muller:2018ibh,
    author = {M\"uller, Berndt and Sch\"afer, Andreas},
    title = "{Chiral magnetic effect and an experimental bound on the late time magnetic field strength}",
    eprint = "1806.10907",
    archivePrefix = "arXiv",
    primaryClass = "hep-ph",
    doi = "10.1103/PhysRevD.98.071902",
    journal = "Phys. Rev. D",
    volume = "98",
    number = "7",
    pages = "071902",
    year = "2018"
}

@article{Sheng:2019kmk,
    author = "Sheng, Xin-Li and Oliva, Lucia and Wang, Qun",
    title = "{What can we learn from the global spin alignment of $\phi$ mesons in heavy-ion collisions?}",
    eprint = "1910.13684",
    archivePrefix = "arXiv",
    primaryClass = "nucl-th",
    doi = "10.1103/PhysRevD.101.096005",
    journal = "Phys. Rev. D",
    volume = "101",
    number = "9",
    pages = "096005",
    year = "2020",
    note = "[Erratum: Phys.Rev.D 105, 099903 (2022)]"
}

@Article{Shi:2019wzi,
  Title                    = {{Signatures of Chiral Magnetic Effect in the Collisions of Isobars}},
  Author                   = {Shi, Shuzhe and Zhang, Hui and Hou, Defu and Liao, Jinfeng},
  Eprint                   = {1910.14010},
  Year                     = {2019},
  Archiveprefix            = {arXiv},
  Primaryclass             = {nucl-th},
  Slaccitation             = {%%CITATION = ARXIV:1910.14010;%%}
}

@Article{Sun:2017xhx,
  Title                    = {{$\Lambda$ hyperon polarization in relativistic heavy ion collisions from a chiral kinetic approach}},
  Author                   = {Sun, Yifeng and Ko, Che Ming},
  DOI                      = {10.1103/PhysRevC.96.024906},
  Eprint                   = {1706.09467},
  Issue                    = {2},
  Month                    = {Aug},
  Number                   = {2},
  Pages                    = {024906},
  URL                      = {https://link.aps.org/doi/10.1103/PhysRevC.96.024906},
  Volume                   = {C96},
  Year                     = {2017},
  Archiveprefix            = {arXiv},
  Journal                  = {Phys. Rev.},
  Numpages                 = {6},
  Primaryclass             = {nucl-th},
  Publisher                = {American Physical Society},
  Slaccitation             = {%%CITATION = ARXIV:1706.09467;%%}
}

@article{Voloshin:2017kqp,
    author = "Voloshin, Sergei A.",
    editor = "Mischke, A. and Kuijer, P.",
    title = "{Vorticity and particle polarization in heavy ion collisions (experimental perspective)}",
    eprint = "1710.08934",
    archivePrefix = "arXiv",
    primaryClass = "nucl-ex",
    doi = "10.1051/epjconf/201817107002",
    journal = "EPJ Web Conf.",
    volume = "171",
    pages = "07002",
    year = "2018"
}

@Article{Xia:2019fjf,
  Title                    = {{Feed-down effect on $\Lambda$ spin polarization}},
  Author                   = {Xia, Xiao-Liang and Li, Hui and Huang, Xu-Guang and Huang, Huan Zhong},
  DOI                      = {10.1103/PhysRevC.100.014913},
  Eprint                   = {1905.03120},
  Number                   = {1},
  Pages                    = {014913},
  Volume                   = {100},
  Year                     = {2019},
  Archiveprefix            = {arXiv},
  Journal                  = {Phys. Rev. C},
  Primaryclass             = {nucl-th}
}

@Article{Xia:2018tes,
  Title                    = {{Probing vorticity structure in heavy-ion collisions by local $\Lambda$ polarization}},
  Author                   = {Xia, Xiao-Liang and Li, Hui and Tang, Ze-Bo and Wang, Qun},
  DOI                      = {10.1103/PhysRevC.98.024905},
  Eprint                   = {1803.00867},
  Pages                    = {024905},
  Volume                   = {98},
  Year                     = {2018},
  Archiveprefix            = {arXiv},
  Journal                  = {Phys. Rev. C},
  Primaryclass             = {nucl-th}
}

@article{Xie:2017upb,
    author = "Xie, Yilong and Wang, Dujuan and Csernai, L\'aszl\'o P.",
    title = "{Global \ensuremath{\Lambda} polarization in high energy collisions}",
    eprint = "1703.03770",
    archivePrefix = "arXiv",
    primaryClass = "nucl-th",
    doi = "10.1103/PhysRevC.95.031901",
    journal = "Phys. Rev. C",
    volume = "95",
    number = "3",
    pages = "031901",
    year = "2017"
}

@Article{Yang:2017sdk,
  Title                    = {{Quark coalescence model for polarized vector mesons and baryons}},
  Author                   = {Yang, Yang-Guang and Fang, Ren-Hong and Wang, Qun and Wang, Xin-Nian},
  DOI                      = {10.1103/PhysRevC.97.034917},
  Eprint                   = {1711.06008},
  Number                   = {3},
  Pages                    = {034917},
  Volume                   = {C97},
  Year                     = {2018},
  Archiveprefix            = {arXiv},
  Journal                  = {Phys. Rev.},
  Primaryclass             = {nucl-th},
  Slaccitation             = {%%CITATION = ARXIV:1711.06008;%%}
}

@article{Fu:2021pok,
    author = "Fu, Baochi and Liu, Shuai Y. F. and Pang, Longgang and Song, Huichao and Yin, Yi",
    title = "{Shear-Induced Spin Polarization in Heavy-Ion Collisions}",
    eprint = "2103.10403",
    archivePrefix = "arXiv",
    primaryClass = "hep-ph",
    doi = "10.1103/PhysRevLett.127.142301",
    journal = "Phys. Rev. Lett.",
    volume = "127",
    number = "14",
    pages = "142301",
    year = "2021"
}

@Article{Fu:2020oxj,
  author        = {Fu, Baochi and Xu, Kai and Huang, Xu-Guang and Song, Huichao},
  journal       = {Phys. Rev. C},
  title         = {{Hydrodynamic study of hyperon spin polarization in relativistic heavy ion collisions}},
  year          = {2021},
  number        = {2},
  pages         = {024903},
  volume        = {103},
  archiveprefix = {arXiv},
  doi           = {10.1103/PhysRevC.103.024903},
  eprint        = {2011.03740},
  primaryclass  = {nucl-th},
}

@article{Becattini:2021iol,
    author = "Becattini, F. and Buzzegoli, M. and Inghirami, G. and Karpenko, I. and Palermo, A.",
    title = "{Local Polarization and Isothermal Local Equilibrium in Relativistic Heavy Ion Collisions}",
    eprint = "2103.14621",
    archivePrefix = "arXiv",
    primaryClass = "nucl-th",
    doi = "10.1103/PhysRevLett.127.272302",
    journal = "Phys. Rev. Lett.",
    volume = "127",
    number = "27",
    pages = "272302",
    year = "2021"
}

@Article{Pang:2012he,
  author        = {Pang, Longgang and Wang, Qun and Wang, Xin-Nian},
  journal       = {Phys. Rev. C},
  title         = {{Effects of initial flow velocity fluctuation in event-by-event (3+1)D hydrodynamics}},
  year          = {2012},
  pages         = {024911},
  volume        = {86},
  archiveprefix = {arXiv},
  doi           = {10.1103/PhysRevC.86.024911},
  eprint        = {1205.5019},
  primaryclass  = {nucl-th},
  reportnumber  = {NT-LBL-12-011},
}

@Article{Pang:2018zzo,
  author        = {Pang, Long-Gang and Petersen, Hannah and Wang, Xin-Nian},
  journal       = {Phys. Rev. C},
  title         = {{Pseudorapidity distribution and decorrelation of anisotropic flow within the open-computing-language implementation CLVisc hydrodynamics}},
  year          = {2018},
  number        = {6},
  pages         = {064918},
  volume        = {97},
  archiveprefix = {arXiv},
  doi           = {10.1103/PhysRevC.97.064918},
  eprint        = {1802.04449},
  primaryclass  = {nucl-th},
}

@Article{Wu:2019eyi,
  author        = {Wu, Hong-Zhong and Pang, Long-Gang and Huang, Xu-Guang and Wang, Qun},
  journal       = {Phys. Rev. Research.},
  title         = {Local spin polarization in high energy heavy ion collisions},
  year          = {2019},
  month         = {Oct},
  pages         = {033058},
  volume        = {1},
  archiveprefix = {arXiv},
  doi           = {10.1103/PhysRevResearch.1.033058},
  eprint        = {1906.09385},
  issue         = {3},
  numpages      = {12},
  primaryclass  = {nucl-th},
  publisher     = {American Physical Society},
  url           = {https://link.aps.org/doi/10.1103/PhysRevResearch.1.033058},
}

@Article{Wu:2020yiz,
  author        = {Wu, Hong-Zhong and Pang, Long-Gang and Huang, Xu-Guang and Wang, Qun},
  journal       = {Nucl. Phys. A},
  title         = {{Local Spin Polarization in 200 GeV Au+Au and 2.76 TeV Pb+Pb Collisions}},
  year          = {2021},
  pages         = {121831},
  volume        = {1005},
  archiveprefix = {arXiv},
  doi           = {10.1016/j.nuclphysa.2020.121831},
  editor        = {Liu, Feng and Wang, Enke and Wang, Xin-Nian and Xu, Nu and Zhang, Ben-Wei},
  eprint        = {2002.03360},
  primaryclass  = {nucl-th},
}

@article{Liu:2021uhn,
	author = "Liu, Shuai Y. F. and Yin, Yi",
	title = "{Spin polarization induced by the hydrodynamic gradients}",
	eprint = "2103.09200",
	archivePrefix = "arXiv",
	primaryClass = "hep-ph",
	doi = "10.1007/JHEP07(2021)188",
	journal = "JHEP",
	volume = "07",
	pages = "188",
	year = "2021"
}

@article{Liu:2020dxg,
	author = "Liu, Shuai Y. F. and Yin, Yi",
	title = "{Spin Hall effect in heavy-ion collisions}",
	eprint = "2006.12421",
	archivePrefix = "arXiv",
	primaryClass = "nucl-th",
	doi = "10.1103/PhysRevD.104.054043",
	journal = "Phys. Rev. D",
	volume = "104",
	number = "5",
	pages = "054043",
	year = "2021"
}

@article{Yi:2021ryh,
	author = "Yi, Cong and Pu, Shi and Yang, Di-Lun",
	title = "{Reexamination of local spin polarization beyond global equilibrium in relativistic heavy ion collisions}",
	eprint = "2106.00238",
	archivePrefix = "arXiv",
	primaryClass = "hep-ph",
	doi = "10.1103/PhysRevC.104.064901",
	journal = "Phys. Rev. C",
	volume = "104",
	number = "6",
	pages = "064901",
	year = "2021"
}

@article{Ryu:2021lnx,
    author = "Ryu, Sangwook and Jupic, Vahidin and Shen, Chun",
    title = "{Probing early-time longitudinal dynamics with the \ensuremath{\Lambda} hyperon's spin polarization in relativistic heavy-ion collisions}",
    eprint = "2106.08125",
    archivePrefix = "arXiv",
    primaryClass = "nucl-th",
    doi = "10.1103/PhysRevC.104.054908",
    journal = "Phys. Rev. C",
    volume = "104",
    number = "5",
    pages = "054908",
    year = "2021"
}

@article{Ambrus:2020oiw,
    author = "Ambrus, Victor E. and Chernodub, M. N.",
    title = "{Hyperon\textendash{}anti-hyperon polarization asymmetry in relativistic heavy-ion collisions as an interplay between chiral and helical vortical effects}",
    eprint = "2010.05831",
    archivePrefix = "arXiv",
    primaryClass = "hep-ph",
    doi = "10.1140/epjc/s10052-022-10002-y",
    journal = "Eur. Phys. J. C",
    volume = "82",
    number = "1",
    pages = "61",
    year = "2022"
}

@Article{Ivanov:2020udj,
  author        = {Ivanov, Yu B.},
  journal       = {Phys. Rev. C},
  title         = {{Global $\Lambda$ polarization in moderately relativistic nuclear collisions}},
  year          = {2021},
  number        = {3},
  pages         = {L031903},
  volume        = {103},
  archiveprefix = {arXiv},
  doi           = {10.1103/PhysRevC.103.L031903},
  eprint        = {2012.07597},
  primaryclass  = {nucl-th},
}

@Article{Guo:2021udq,
  author        = {Guo, Yu and Liao, Jinfeng and Wang, Enke and Xing, Hongxi and Zhang, Hui},
  journal       = {Phys. Rev. C},
  title         = {{Hyperon polarization from the vortical fluid in low-energy nuclear collisions}},
  year          = {2021},
  number        = {4},
  pages         = {L041902},
  volume        = {104},
  archiveprefix = {arXiv},
  doi           = {10.1103/PhysRevC.104.L041902},
  eprint        = {2105.13481},
  primaryclass  = {nucl-th},
}

@Article{Deng:2020ygd,
  author        = {Deng, Xian-Gai and Huang, Xu-Guang and Ma, Yu-Gang and Zhang, Song},
  journal       = {Phys. Rev. C},
  title         = {{Vorticity in low-energy heavy-ion collisions}},
  year          = {2020},
  number        = {6},
  pages         = {064908},
  volume        = {101},
  archiveprefix = {arXiv},
  doi           = {10.1103/PhysRevC.101.064908},
  eprint        = {2001.01371},
  primaryclass  = {nucl-th},
}

@article{Deng:2021miw,
    author = "Deng, Xian-Gai and Huang, Xu-Guang and Ma, Yu-Gang",
    title = "{Lambda polarization in 108Ag+108Ag and 197Au+197Au collisions around a few GeV}",
    eprint = "2109.09956",
    archivePrefix = "arXiv",
    primaryClass = "nucl-th",
    doi = "10.1016/j.physletb.2022.137560",
    journal = "Phys. Lett. B",
    volume = "835",
    pages = "137560",
    year = "2022"
}

@article{STAR:2021beb,
    author = "Abdallah, M. S. and others",
    collaboration = "STAR",
    title = "{Global $\Lambda$-hyperon polarization in Au+Au collisions at $\sqrt {s_{NN}}$=3~GeV}",
    eprint = "2108.00044",
    archivePrefix = "arXiv",
    primaryClass = "nucl-ex",
    doi = "10.1103/PhysRevC.104.L061901",
    journal = "Phys. Rev. C",
    volume = "104",
    number = "6",
    pages = "L061901",
    year = "2021"
}

@Article{Lei:2021mvp,
  author        = {Lei, Anke and Wang, Dujuan and zhou, Dai-Mei and Sa, Ben-Hao and Csernai, Laszlo Pal},
  journal       = {Phys. Rev. C},
  title         = {{Vorticity and \ensuremath{\Lambda} polarization in the microscopic transport model PACIAE}},
  year          = {2021},
  number        = {5},
  pages         = {054903},
  volume        = {104},
  archiveprefix = {arXiv},
  doi           = {10.1103/PhysRevC.104.054903},
  eprint        = {2110.13485},
  primaryclass  = {nucl-th},
}

@article{Ayala:2021xrn,
    author = "Ayala, Alejandro and Dom\'\i{}nguez, Isabel and Maldonado, Ivonne and Tejeda-Yeomans, Mar\'\i{}a Elena",
    title = "{Rise and fall of \ensuremath{\Lambda} and \ensuremath{\Lambda}\textasciimacron{} global polarization in semi-central heavy-ion collisions at HADES, NICA and RHIC energies from the core-corona model}",
    eprint = "2106.14379",
    archivePrefix = "arXiv",
    primaryClass = "hep-ph",
    doi = "10.1103/PhysRevC.105.034907",
    journal = "Phys. Rev. C",
    volume = "105",
    number = "3",
    pages = "034907",
    year = "2022"
}

@InProceedings{Li:2021jvn,
  author        = {Li, Hui and Xia, Xiao-Liang and Huang, Xu-Guang and Huang, Huan Zhong},
  booktitle     = {{19th International Conference on Strangeness in Quark Matter}},
  title         = {{Global hyperon polarization and effects of decay feeddown}},
  year          = {2021},
  month         = {8},
  archiveprefix = {arXiv},
  eprint        = {2108.04111},
  primaryclass  = {nucl-th},
}

@article{Sheng:2022wsy,
    author = "Sheng, Xin-Li and Oliva, Lucia and Liang, Zuo-Tang and Wang, Qun and Wang, Xin-Nian",
    title = "{Spin Alignment of Vector Mesons in Heavy-Ion Collisions}",
    eprint = "2205.15689",
    archivePrefix = "arXiv",
    primaryClass = "nucl-th",
    reportNumber = "USTC-ICTS/PCFT-22-16",
    doi = "10.1103/PhysRevLett.131.042304",
    journal = "Phys. Rev. Lett.",
    volume = "131",
    number = "4",
    pages = "042304",
    year = "2023"
}

@article{Sheng:2022ffb,
    author = "Sheng, Xin-Li and Oliva, Lucia and Liang, Zuo-Tang and Wang, Qun and Wang, Xin-Nian",
    title = "{Relativistic spin dynamics for vector mesons}",
    eprint = "2206.05868",
    archivePrefix = "arXiv",
    primaryClass = "hep-ph",
    reportNumber = "USTC-ICTS/PCFT-22-18",
    doi = "10.1103/PhysRevD.109.036004",
    journal = "Phys. Rev. D",
    volume = "109",
    number = "3",
    pages = "036004",
    year = "2024"
}

@Article{Becattini:2016gvu,
  author        = {Becattini, F. and Karpenko, I. and Lisa, M. and Upsal, I. and Voloshin, S.},
  journal       = {Phys. Rev. C},
  title         = {{Global hyperon polarization at local thermodynamic equilibrium with vorticity, magnetic field and feed-down}},
  year          = {2017},
  number        = {5},
  pages         = {054902},
  volume        = {95},
  archiveprefix = {arXiv},
  doi           = {10.1103/PhysRevC.95.054902},
  eprint        = {1610.02506},
  primaryclass  = {nucl-th},
}

@Article{Monnai:2019hkn,
  author        = {Monnai, Akihiko and Schenke, Bj\"orn and Shen, Chun},
  journal       = {Phys. Rev. C},
  title         = {{Equation of state at finite densities for QCD matter in nuclear collisions}},
  year          = {2019},
  number        = {2},
  pages         = {024907},
  volume        = {100},
  archiveprefix = {arXiv},
  doi           = {10.1103/PhysRevC.100.024907},
  eprint        = {1902.05095},
  primaryclass  = {nucl-th},
  reportnumber  = {KEK-TH-2106},
}

@Article{Monnai:2021kgu,
  author        = {Monnai, Akihiko and Schenke, Bj\"orn and Shen, Chun},
  journal       = {Int. J. Mod. Phys. A},
  title         = {{QCD Equation of State at Finite Chemical Potentials for Relativistic Nuclear Collisions}},
  year          = {2021},
  number        = {07},
  pages         = {2130007},
  volume        = {36},
  archiveprefix = {arXiv},
  doi           = {10.1142/S0217751X21300076},
  eprint        = {2101.11591},
  primaryclass  = {nucl-th},
}

@Article{Wu:2021fjf,
  author        = {Wu, Xiang-Yu and Qin, Guang-You and Pang, Long-Gang and Wang, Xin-Nian},
  journal       = {Phys. Rev. C},
  title         = {{(3+1)-D viscous hydrodynamics at finite net baryon density: Identified particle spectra, anisotropic flows, and flow fluctuations across energies relevant to the beam-energy scan at RHIC}},
  year          = {2022},
  number        = {3},
  pages         = {034909},
  volume        = {105},
  archiveprefix = {arXiv},
  doi           = {10.1103/PhysRevC.105.034909},
  eprint        = {2107.04949},
  primaryclass  = {hep-ph},
}

@Article{Weil:2016zrk,
  author        = {Weil, J. and others},
  journal       = {Phys. Rev. C},
  title         = {{Particle production and equilibrium properties within a new hadron transport approach for heavy-ion collisions}},
  year          = {2016},
  number        = {5},
  pages         = {054905},
  volume        = {94},
  archiveprefix = {arXiv},
  doi           = {10.1103/PhysRevC.94.054905},
  eprint        = {1606.06642},
  primaryclass  = {nucl-th},
}

@Article{Schafer:2019edr,
  author        = {Sch\"afer, Anna and Torres-Rincon, Juan M. and Rothermel, Jonas and Ehlert, Niklas and Gale, Charles and Elfner, Hannah},
  journal       = {Phys. Rev. D},
  title         = {{Benchmarking a nonequilibrium approach to photon emission in relativistic heavy-ion collisions}},
  year          = {2019},
  number        = {11},
  pages         = {114021},
  volume        = {99},
  archiveprefix = {arXiv},
  doi           = {10.1103/PhysRevD.99.114021},
  eprint        = {1902.07564},
  primaryclass  = {nucl-th},
}

@Article{Mohs:2019iee,
  author        = {Mohs, Justin and Ryu, Sangwook and Elfner, Hannah},
  journal       = {J. Phys. G},
  title         = {{Particle Production via Strings and Baryon Stopping within a Hadronic Transport Approach}},
  year          = {2020},
  number        = {6},
  pages         = {065101},
  volume        = {47},
  archiveprefix = {arXiv},
  doi           = {10.1088/1361-6471/ab7bd1},
  eprint        = {1909.05586},
  primaryclass  = {nucl-th},
}

@Article{Hammelmann:2019vwd,
  author        = {Hammelmann, Jan and Soto-Ontoso, Alba and Alvioli, Massimiliano and Elfner, Hannah and Strikman, Mark},
  journal       = {Phys. Rev. C},
  title         = {{Influence of the neutron-skin effect on nuclear isobar collisions at energies available at the BNL Relativistic Heavy Ion Collider}},
  year          = {2020},
  number        = {6},
  pages         = {061901},
  volume        = {101},
  archiveprefix = {arXiv},
  doi           = {10.1103/PhysRevC.101.061901},
  eprint        = {1908.10231},
  primaryclass  = {nucl-th},
}

@Article{Mohs:2020awg,
  author        = {Mohs, Justin and Ege, Martha and Elfner, Hannah and Mayer, Markus},
  journal       = {Phys. Rev. C},
  title         = {{Collective flow at SIS energies within a hadronic transport approach: Influence of light nuclei formation and equation~of state}},
  year          = {2022},
  number        = {3},
  pages         = {034906},
  volume        = {105},
  archiveprefix = {arXiv},
  doi           = {10.1103/PhysRevC.105.034906},
  eprint        = {2012.11454},
  primaryclass  = {nucl-th},
}

@article{Schafer:2021csj,
    author = {Sch\"afer, Anna and Karpenko, Iurii and Wu, Xiang-Yu and Hammelmann, Jan and Elfner, Hannah},
    collaboration = "SMASH",
    title = "{Particle production in a hybrid approach for a beam energy scan of Au+Au/Pb+Pb collisions between $\sqrt{s_\textrm{NN}}$ = 4.3 GeV and $\sqrt{s_\textrm{NN}}$ = 200.0 GeV}",
    eprint = "2112.08724",
    archivePrefix = "arXiv",
    primaryClass = "hep-ph",
    doi = "10.1140/epja/s10050-022-00872-x",
    journal = "Eur. Phys. J. A",
    volume = "58",
    number = "11",
    pages = "230",
    year = "2022"
}

@Article{Inghirami:2022afu,
  author        = {Inghirami, Gabriele and Elfner, Hannah},
  journal       = {Eur. Phys. J. C},
  title         = {{The applicability of hydrodynamics in heavy ion collisions at $\sqrt{s_\mathrm{NN}}$~=~2.4\textendash{}7.7~GeV}},
  year          = {2022},
  number        = {9},
  pages         = {796},
  volume        = {82},
  archiveprefix = {arXiv},
  doi           = {10.1140/epjc/s10052-022-10718-x},
  eprint        = {2201.05934},
  primaryclass  = {hep-ph},
}

@Article{Xia:2020tyd,
  author        = {Xia, Xiao-Liang and Li, Hui and Huang, Xu-Guang and Zhong Huang, Huan},
  journal       = {Phys. Lett. B},
  title         = {{Local spin alignment of vector mesons in relativistic heavy-ion collisions}},
  year          = {2021},
  pages         = {136325},
  volume        = {817},
  archiveprefix = {arXiv},
  doi           = {10.1016/j.physletb.2021.136325},
  eprint        = {2010.01474},
  primaryclass  = {nucl-th},
}

@Article{Liang:2004ph,
  author        = {Liang, Zuo-Tang and Wang, Xin-Nian},
  journal       = {Phys. Rev. Lett.},
  title         = {{Globally polarized quark-gluon plasma in non-central A+A collisions}},
  year          = {2005},
  note          = {[Erratum: Phys.Rev.Lett. 96, 039901 (2006)]},
  pages         = {102301},
  volume        = {94},
  archiveprefix = {arXiv},
  doi           = {10.1103/PhysRevLett.94.102301},
  eprint        = {nucl-th/0410079},
  reportnumber  = {LBNL-56383},
}

@Article{Liang:2004xn,
  author        = {Liang, Zuo-Tang and Wang, Xin-Nian},
  journal       = {Phys. Lett. B},
  title         = {{Spin alignment of vector mesons in non-central A+A collisions}},
  year          = {2005},
  pages         = {20--26},
  volume        = {629},
  archiveprefix = {arXiv},
  doi           = {10.1016/j.physletb.2005.09.060},
  eprint        = {nucl-th/0411101},
  reportnumber  = {LBNL-56659},
}

@Article{Li:2017slc,
  author        = {Li, Hui and Pang, Long-Gang and Wang, Qun and Xia, Xiao-Liang},
  journal       = {Phys. Rev. C},
  title         = {{Global $\Lambda$ polarization in heavy-ion collisions from a transport model}},
  year          = {2017},
  number        = {5},
  pages         = {054908},
  volume        = {96},
  archiveprefix = {arXiv},
  doi           = {10.1103/PhysRevC.96.054908},
  eprint        = {1704.01507},
  primaryclass  = {nucl-th},
}

@Article{Vitiuk:2019rfv,
  author        = {Vitiuk, O. and Bravina, L. V. and Zabrodin, E. E.},
  journal       = {Phys. Lett. B},
  title         = {{Is different $\Lambda$ and $\bar \Lambda$ polarization caused by different spatio-temporal freeze-out picture?}},
  year          = {2020},
  pages         = {135298},
  volume        = {803},
  archiveprefix = {arXiv},
  doi           = {10.1016/j.physletb.2020.135298},
  eprint        = {1910.06292},
  primaryclass  = {hep-ph},
}

@Article{Fang:2016vpj,
  author        = {Fang, Ren-hong and Pang, Long-gang and Wang, Qun and Wang, Xin-nian},
  journal       = {Phys. Rev. C},
  title         = {{Polarization of massive fermions in a vortical fluid}},
  year          = {2016},
  number        = {2},
  pages         = {024904},
  volume        = {94},
  archiveprefix = {arXiv},
  doi           = {10.1103/PhysRevC.94.024904},
  eprint        = {1604.04036},
  primaryclass  = {nucl-th},
  reportnumber  = {ICTS-USTC-16-05},
}

@Article{Peng:2022cya,
  author        = {Peng, Hao-Hao and Wu, Sihao and Wang, Ren-jie and She, Duan and Pu, Shi},
  title         = {{Anomalous magnetohydrodynamics with temperature-dependent electric conductivity and application to the global polarization}},
  year          = {2022},
  month         = {11},
  archiveprefix = {arXiv},
  eprint        = {2211.11286},
  primaryclass  = {hep-ph},
}

@Article{STAR:2023eck,
  author        = {Abdulhamid, Muhammad and others},
  journal       = {Phys. Rev. Lett.},
  title         = {{Hyperon Polarization along the Beam Direction Relative to the Second and Third Harmonic Event Planes in Isobar Collisions at $\sqrt{s_\mathrm{NN}}$=200\,\,GeV}},
  year          = {2023},
  number        = {20},
  pages         = {202301},
  volume        = {131},
  archiveprefix = {arXiv},
  collaboration = {STAR},
  doi           = {10.1103/PhysRevLett.131.202301},
  eprint        = {2303.09074},
  primaryclass  = {nucl-ex},
}

@Article{ALICE:2021pzu,
  author        = {Acharya, Shreyasi and others},
  journal       = {Phys. Rev. Lett.},
  title         = {{Polarization of $\Lambda$ and $\bar \Lambda$ Hyperons along the Beam Direction in Pb-Pb Collisions at $\sqrt {s_{NN}}$=5.02\,\,TeV}},
  year          = {2022},
  number        = {17},
  pages         = {172005},
  volume        = {128},
  archiveprefix = {arXiv},
  collaboration = {ALICE},
  doi           = {10.1103/PhysRevLett.128.172005},
  eprint        = {2107.11183},
  primaryclass  = {nucl-ex},
  reportnumber  = {CERN-EP-2021-148},
}

@article{Yi:2023tgg,
    author = "Yi, Cong and Wu, Xiang-Yu and Yang, Di-Lun and Gao, Jian-Hua and Pu, Shi and Qin, Guang-You",
    title = "{Probing vortical structures in heavy-ion collisions at RHIC-BES energies through helicity polarization}",
    eprint = "2304.08777",
    archivePrefix = "arXiv",
    primaryClass = "hep-ph",
    doi = "10.1103/PhysRevC.109.L011901",
    journal = "Phys. Rev. C",
    volume = "109",
    number = "1",
    pages = "L011901",
    year = "2024"
}

@article{Sheng:2022ssp,
    author = "Sheng, Xin-Li and Yang, Shu-Yun and Zou, Yao-Lin and Hou, Defu",
    title = "{Mass splitting and spin alignment for $\phi $ mesons in a magnetic field in NJL model}",
    eprint = "2209.01872",
    archivePrefix = "arXiv",
    primaryClass = "nucl-th",
    doi = "10.1140/epjc/s10052-024-12643-7",
    journal = "Eur. Phys. J. C",
    volume = "84",
    number = "3",
    pages = "299",
    year = "2024"
}

@Article{Li:2022vmb,
  author        = {Li, Feng and Liu, Shuai Y. F.},
  title         = {{Tensor Polarization and Spectral Properties of Vector Meson in QCD Medium}},
  year          = {2022},
  month         = {6},
  archiveprefix = {arXiv},
  eprint        = {2206.11890},
  primaryclass  = {nucl-th},
}

@article{Sheng:2023urn,
    author = "Sheng, Xin-Li and Pu, Shi and Wang, Qun",
    title = "{Momentum dependence of the spin alignment of the \ensuremath{\phi} meson}",
    eprint = "2308.14038",
    archivePrefix = "arXiv",
    primaryClass = "nucl-th",
    doi = "10.1103/PhysRevC.108.054902",
    journal = "Phys. Rev. C",
    volume = "108",
    number = "5",
    pages = "054902",
    year = "2023"
}

@article{Dong:2023cng,
    author = "Dong, Wen-Bo and Yin, Yi-Liang and Sheng, Xin-Li and Yang, Shi-Zheng and Wang, Qun",
    title = "{Linear response theory for spin alignment of vector mesons in thermal media}",
    eprint = "2311.18400",
    archivePrefix = "arXiv",
    primaryClass = "hep-ph",
    doi = "10.1103/PhysRevD.109.056025",
    journal = "Phys. Rev. D",
    volume = "109",
    number = "5",
    pages = "056025",
    year = "2024"
}

@Article{Wagner:2022gza,
  author        = {Wagner, David and Weickgenannt, Nora and Speranza, Enrico},
  journal       = {Phys. Rev. Res.},
  title         = {{Generating tensor polarization from shear stress}},
  year          = {2023},
  number        = {1},
  pages         = {013187},
  volume        = {5},
  archiveprefix = {arXiv},
  doi           = {10.1103/PhysRevResearch.5.013187},
  eprint        = {2207.01111},
  primaryclass  = {nucl-th},
}

@article{Sheng:2024kgg,
    author = "Sheng, Xin-Li and Zhao, Yan-Qing and Li, Si-Wen and Becattini, Francesco and Hou, Defu",
    title = "{Holographic spin alignment for vector mesons}",
    eprint = "2403.07522",
    archivePrefix = "arXiv",
    primaryClass = "hep-ph",
    doi = "10.1103/PhysRevD.110.056047",
    journal = "Phys. Rev. D",
    volume = "110",
    number = "5",
    pages = "056047",
    year = "2024"
}

@article{Lv:2024uev,
    author = "Lv, Ji-peng and Yu, Zi-han and Liang, Zuo-tang and Wang, Qun and Wang, Xin-Nian",
    title = "{Global quark spin correlations in relativistic heavy ion collisions}",
    eprint = "2402.13721",
    archivePrefix = "arXiv",
    primaryClass = "hep-ph",
    doi = "10.1103/PhysRevD.109.114003",
    journal = "Phys. Rev. D",
    volume = "109",
    number = "11",
    pages = "114003",
    year = "2024"
}

@Article{Becattini:2024uha,
  author        = {Becattini, Francesco and Buzzegoli, Matteo and Niida, Takafumi and Pu, Shi and Tang, Ai-Hong and Wang, Qun},
  title         = {{Spin polarization in relativistic heavy-ion collisions}},
  year          = {2024},
  month         = {2},
  archiveprefix = {arXiv},
  eprint        = {2402.04540},
  primaryclass  = {nucl-th},
}

\end{document}